\documentclass[preprint,superscriptaddress,aps]{revtex4}

\usepackage{amsmath} 
\usepackage{amssymb} 
\usepackage{amsfonts}

\usepackage[pdfencoding=auto]{hyperref}

\usepackage{graphicx} 
\usepackage{color,soul}

\usepackage{array}
\usepackage{multirow}

\begin{document}

\bibliographystyle{apsrev4-1} 

\title{Solvent fluctuations induce non-Markovian kinetics in hydrophobic pocket-ligand binding} 

\author{R. Gregor Wei\ss}
\affiliation{Institut f{\"u}r Physik, Humboldt-Universit{\"a}t zu Berlin, Newtonstr.~15, D-12489 Berlin, Germany}
\affiliation{Institut f{\"u}r Weiche Materie and Funktionale Materialen, Helmholtz-Zentrum Berlin, Hahn-Meitner Platz 1, D-14109 Berlin, Germany}
\author{Piotr Setny}
\affiliation{Centre of New Technologies, University of Warsaw, 00-927 Warsaw, Poland}
\author{Joachim Dzubiella}
\thanks{To whom correspondence should be addressed. E-mail: joachim.dzubiella@helmholtz-berlin.de}
\affiliation{Institut f{\"u}r Physik, Humboldt-Universit{\"a}t zu Berlin, Newtonstr.~15, D-12489 Berlin, Germany}
\affiliation{Institut f{\"u}r Weiche Materie and Funktionale Materialen, Helmholtz-Zentrum Berlin, Hahn-Meitner Platz 1, D-14109 Berlin, Germany}

\begin{abstract}
We investigate the impact of water fluctuations on the key-lock association kinetics of a hydrophobic ligand (key) binding to a
hydrophobic pocket (lock) by means of a minimalistic stochastic model system. It
describes the collective hydration behavior 
of the pocket by bimodal fluctuations of a water-pocket interface that dynamically couples to the diffusive
motion of the approaching ligand via the hydrophobic interaction. This leads to a set
of overdamped Langevin equations in 2D-coordinate-space, 
that is Markovian in each dimension. Numerical simulations demonstrate locally increased friction of the ligand, 
decelerated binding kinetics, and local non-Markovian (memory) effects in the ligand's reaction coordinate as found previously in
explicit-water molecular dynamics studies of model hydrophobic pocket-ligand binding~\cite{Setny:PNAS,Mondal:PNAS}. 
Our minimalistic model elucidates the origin of effectively enhanced friction in the process that can be traced back 
to long-time decays in the force-autocorrelation function induced by the effective, spatially fluctuating pocket-ligand interaction.
Furthermore, we construct a generalized 1D-Langevin description including a spatially local memory function that  enables
further interpretation and a semi-analytical quantification of the results of the coupled 2D-system.

\end{abstract}

\maketitle

\section{Introduction}
Nature expresses a strong versatility in its creation of substrates as ligands and binding
sites as receptors, thereby utilizing the complex properties of water as natural solvent environment. 
This evolutionary framework facilitates multifaceted kinetics of biomolecular recognition and association 
and  has led to substantial interdisciplinary research in the last decades towards
fundamentally comprehending the natural mechanisms of ligand-receptor (or key-lock) 
binding as a  part of life's cycle. Consequently, one substantive objective that recurs eminently 
in science is the detailed molecular understanding of ligand binding processes for the design and 
development of pharmaceutical substances.  

Many experimental as well as theoretical studies on the thermodynamics
\cite{Gellman,Gilson&Zhou,Woo&Roux,Wang&Wang,Head&Marshall} of
an increasing number of ligand-enzyme complexes have provided insight 
about binding free energies, namely binding affinities, of the individual systems. 
Taken alone, however, thermodynamics cannot predict exact kinetic properties. 
Yet rates of binding and unbinding events are crucial factors determining
drugs efficiency~\cite{McCammon,Pan&Shaw,Tiwary&Parrinello}.  

Pioneering research recognized dynamic couplings as important
component for estimating the time scale of molecular  biological processes
\cite{Beece&Yue,Grote&Hynes,Hänggi,Doster,Frauenfelder&Wolynes}.
Beece~\textit{et~al.} \cite{Beece&Yue} discussed the impact of
structural fluctuations of enzymes on the migration kinetics of
substrates. They observed large effects of protein fluctuations on
binding rates in the thoroughly explored process of carbon monoxide migration
to myoglobin. Therein crucial impact roots from fluctuations of
opening-closing conformations of protein channels. Along their
line of arguments, different solvent viscosities, thus different environments to the ligand-enzyme
complex, change the protein's internal fluctuations which couple to
the kinetics of ligand migration. In general, internal
barriers of conformational fluctuations in a protein can be
comparable to the thermal energy~\cite{Batista&Robert}, facilitating
time scales to be similar to those of ligand kinetics~\cite{Ishikawa&Fayer,Bernardi&Schulten}.

Specific work on inactive-active, e.g., open-closed, conformational
transitions of biomolecular receptors observed and proposed kinetic models by the
\textit{induced fit} and \textit{conformational selection}
paradigms~\cite{Zhou,Cai&Zhou}. Within these models the
conformational transitions are treated as distinct states taken
with given probability and transition rates fulfilling detailed
balance. Hence the ratios of transition rates and
state-probabilities determine whether ligand migration
\textit{induces} the active conformation for binding or whether
binding occurs predominantly when the pocket \textit{conformation}
is active long before ligand association. Extending this picture,
Zhou and co-workers~\cite{Zhou,Cai&Zhou} allow coupling of the
conformational kinetics to ligand migration whereas they
utilize Markovian kinetics in a two state model.

A more general discussion~\cite{Grote&Hynes,Hänggi,Doster}
describes dynamic coupling of substrates and enzymes by an
underdamped kinetic description. It models ligand migration by a
generalized Langevin equation (GLE) including memory on random
velocity changes. The time scales of the memory kernel are incorporated
in an additional multiplication factor to conventional
transition state theory for rate calculation
over a barrier. Hence calculations for individual
ligand-pocket systems estimate relative retardation or
acceleration to a Markovian crossing rate. This extension to
reaction rate theory is also known as Grote-Hynes theory~\cite{Grote&Hynes}.

Direct coupling of water dynamics to hydrophobic key-lock
binding kinetics was recently observed by
Setny~\textit{et~al.}~\cite{Setny:PNAS}. By means of explicit-water
molecular dynamics (MD) simulations of a model hydrophobic pocket-ligand system 
they found long-time correlation effects, i.e., the hint to memory, in position
and force correlations when the ligand was situated
in the immediate vicinity  of the pocket. It was argued that their origin were 
pocket water occupancy fluctuations that occur due to capillary evaporation in the small
confinement between hydrophobic ligand and hydrophobic pocket~\cite{Setny:PRL}, 
yielding \textit{dry} states, without water inside the pocket, and
\textit{wet} states, with a maximally solvated pocket. Also the ligand position was
shown to sensitively affect the bimodal (dry-wet) pocket hydration distribution and time
scale.
A Markovian
description of mean binding times utilizing potential of mean
force and a spatially dependent friction could not reproduce
binding times directly calculated in the MD simulation. Hence, a 
non-Markovian treatment of ligand migration within hydrophobic key-lock 
binding processes was proposed.  In a subsequent study by Mondal~\textit{et~al.}~\cite{Mondal:PNAS} 
on MD simulations of a very related pocket-ligand model system  the two hydration states were 
utilized in a reaction-diffusion model similar to the
descriptions for inactive-active confirmation kinetics~~\cite{Zhou,Cai&Zhou}. This two-state model  
then improved the rate predictions from MD simulations~\cite{Mondal:PNAS}.

In this work we introduce a minimalistic stochastic model for the kinetic binding of a ligand
to a hydrophobic pocket that exhibits bimodal wet-dry transitions.
Here, one stochastic
coordinate is a bimodally fluctuating pocket-water interface and the second the position of a ligand that 
travels in one spatial dimension. Both coordinates are coupled via a hydrophobic interaction between 
ligand and the fluctuating interface.  Mathematically, we describe that by two coupled Langevin equations.  
In Section \ref{sec:model} the details of the model are described. 
Section \ref{sec:numerical} presents numerical evaluation of the model system analyzing ligand 
binding kinetics. Comparison of mean binding times from numerical simulation to a
corresponding memoryless stochastic process demonstrates the break 
down of Markovian behavior for the single reaction coordinate of
the ligand. Friction calculations indicate that additional damping in hydrophobic key-lock 
association originates from the fluctuating potential on slow time scales.
In this way, numeric evaluations tightly follow the procedure in
Ref.~\cite{Setny:PNAS} answering the previously open questions emphasized by
similar findings with the minimalistic model here. To further
corroborate these findings Section \ref{sec:GLE} deals with a complementary theory describing an effective 1D-reaction
coordinate ligand system in terms of a generalized Langevin equation including a local 
memory function. This formulation enables further interpretation and a semi-analytical quantification 
of the results of the overdamped but coupled 2D-reaction coordinate system 
from Section \ref{sec:friction}. We conclude our study in Section \ref{sec:conclusion}.

\section{Langevin model}

\label{sec:model}

\begin{figure}[t]
\centerline{\includegraphics[width=8.6cm]{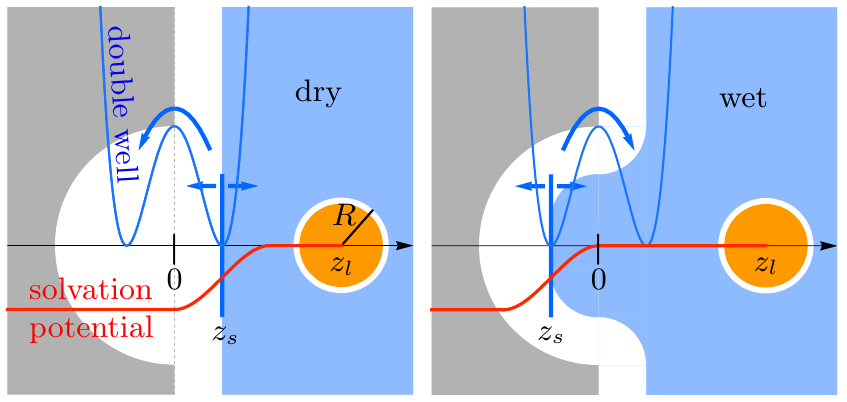}}
\caption{Illustration of the stochastic model: The interface as
pseudo-particle (thick blue 'surface' line) at position $z_s$ fluctuates in the
pocket in a
double-well potential (eq.~\eqref{eq1}, blue curve) in $z$-direction leaving the pocket either dehydrated/dry (left) or
hydrated/wet (right). The ligand (orange circle) at position $z_l$ diffuses freely in one-dimension on the $z$-axis
perpendicular to the pocketed wall. It interacts with the interface by an
attractive interaction potential (eq.~\eqref{eq3}, red line). As illustrated here the interaction potential moves with interface.
Note the model is spatially purely one-dimensional and the schematics 
of the pocketed wall (gray) and bulk water area (blue) are just shown for illustration.}
\label{fig1}
\end{figure}

Our minimalistic stochastic model assumes that the ligand is a particle
diffusing in one spatial direction $z$ driven by a stochastic random force. The
surrounding water creates a liquid-vapor interface near the hydrophobic walls of
the pocket. Water, and thus the interface, can penetrate the pocket leaving it in a 'wet'-state or
in a 'dry'-state, if it resides in front of the pocket. This behavior is met by a
pseudo-particle that effectively describes interface motion in the pocket region
around $z=0$.  
A schematic setup of the interface-ligand system is illustrated in
Fig.~\ref{fig1}. It shows the pseudo-particle as a thick blue line representing a
sharp water-vapor surface at $z_s$ and an orange spherical ligand of radius $R$
at $z_l$. 

The ligand diffuses with the properties of a spherical particle in water utilizing
the Einstein relation $D = k_{\rm B} T / 6 \pi \eta R$, where $k_{\rm B}$ is the Boltzmann
constant,  $\eta$ the viscosity of the solvent and $T$ the temperature. 
Dynamic coupling may occur when interface fluctuations and ligand
kinetics are on a similar time scale. Hence, simply choosing equal
diffusivity for both ligand and pseudo-particles facilitates the condition having
both time scales with comparable magnitude.
In our model the energy scale $\mathrm{k_BT}$ and the length scale $R$ 
are set to one in the following, as well as the diffusion constant $D$. 
This introduces a Brownian length scale $\lambda_B=R$ and time scale
$\tau_B=\lambda_B^2/D$. 
A detailed comment on the relation of units and physical constants to the 
previous explicit-water MD simulations is provided in the Appendix~\ref{SI:units}. 

As motivated from previous MD studies, we assume that the interface fluctuates
bimodally between positions inside or outside the pocket. This models enhanced
fluctuations of the water interface penetrating into the pocket. Thus the interface
moves as a Brownian pseudo-particle in an external double-well
potential 
\begin{equation} 
V_{dw}(z_s) = \frac{h}{\lambda_B^4} \left ( z_s^2 - \lambda_B^2 \right )^2 + b\cdot z_s 
\label{eq1} 
\end{equation} 
which is drawn as blue
curve in Fig. \ref{fig1}. For $b=0$, the positions of the two
wells are situated at $\pm \lambda_B$, 
and $h$, in our energy units, is the height of the barrier which lies at $z=0$. To
further enable
changes in relative depths of 'dry' and 'wet' wells we introduce a
bias given by the linearity constant $b$ in $\mathrm{k_BT}/\lambda_B$. 

\begin{figure}[t]
\centerline{\includegraphics[width=8.6cm]{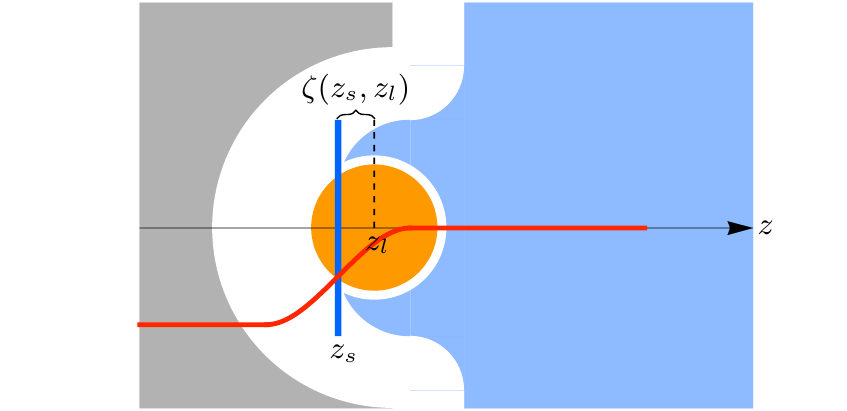}}
\caption{If the distance $\zeta(z_s,z_l) = z_l - z_s$ fulfills $\Theta(R-|\zeta|)=1$ the water-vapor interface (vertical blue
line) and the ligand (orange sphere) interact by the \textit{solvation potential}
(eq.~\eqref{eq3}, red curve). The interaction scales by the
solvated volume of the ligand, which is the portion on the right hand side of the interface.}
\label{fig2}
\end{figure}

A pair potential acting between the
interface and the ligand 
accounts for energetic contribution of solvation as the ligand
passes through the water interface (see Fig.~\ref{fig2}). The resulting
\textit{solvation potential} is designed such that it pushes
the ligand out of the solvent into
the pocket ($z_l < 0$). At the same time, following the principle
of action-reaction, the interaction pulls the interfacial water out of the
pocket ($z_s > 0$), which conceptually corresponds to ligand-induced drying
transition. For
small solutes solvation energy approximately scales linearly with
solvent excluded volume $\Delta G \propto V$, whereas after the transition at a
crossover length-scale $l_{c}$ it is proportional to solvent accessible
surface area $A$, $\Delta G = \gamma \cdot A$
with $\gamma$ as surface tension \cite{Chandler:Nature}. Modeling
microscopic key-lock binding with a small-sized ligand, we choose the
solvation potential to scale linearly with solute volume, or solvent excluded
volume. We
demand a
reasonable proportionality constant $\Gamma$ to fulfill $\Delta G
(l_{c}) = \Gamma \cdot V(l_c) \equiv \gamma \cdot A(l_c)$ at the crossover
length-scale, which thus yields
\begin{equation}
\Gamma = 3 \gamma / l_c \,.
\label{eq2}
\end{equation}
For pure water surface tension we calculate
$\Gamma=2.95~\mathrm{k_BT}/\lambda_B^3$ (see Appendix~\ref{SI:units})
such that the effective solvation energy is roughly
$12.36~\mathrm{k_BT}$, which is comparable to the results of explicit water
simulations \cite{Setny:JCTC:2010}. Solvent excluded volume changes with ligand
distance to the water interface, $\zeta (z_s,z_l) = z_l - z_s$, as it is
illustrated in Fig.~2. 
The solvation
potential is then written as
\begin{equation}
\Delta G(z_s,z_l) = \Gamma \left [ \frac{4\pi}{3} R^3 - \frac{\pi}{3} (R-\zeta)^2 (2R+\zeta) \right ]
\label{eq3}
\end{equation}
which gives a parabolic pair force, acting on particle $x = s,l$ (solvent or ligand) 
\begin{equation}
F_{sol}^{x}(z_s, z_l) = - \frac{\mathrm{d} \Delta G(\zeta) }{\mathrm{d} \, \zeta} \frac{\mathrm{d} \zeta}{\mathrm{d} z_x} = \pi
\Gamma (\zeta^2 - R^2) \frac{\mathrm{d} \zeta}{\mathrm{d} z_x}
\label{eq4}
\end{equation}
The solvation potential only acts when the separation of the ligand to the
interface is smaller than $R$ which can be expressed by the Heaviside
step function $\Theta(R-|\zeta|)$. For clarity, however, it is omitted
in eq.~\eqref{eq3} and \eqref{eq4}.

The gray wall in Fig.~\ref{fig1} embedding the cavity is only drawn
representatively. Naturally the system describes the bimodal water
interface fluctuations due to hydrophobic confinement, but a potential
incorporating steric repulsion and van
der Waals attraction is omitted. It is not needed here since numerical
simulation of
unrestrained ligand motion is aborted every time the ligand is bound,
namely when $z_l=-1.25$. 

Also, we fix a reflective boundary to a given distance $z_{max}$ to the pocket in
order to avoid the ligand diffusing far away from the pocket. Throughout the main
body of the paper $z_{max}=5$, whereas in Appendix~\ref{SI:size} the impact of the
choice of $z_{max}$ is discussed.

In summary, two nonlinearly coupled Langevin equations describe the key-lock system by
\begin{widetext}
	\begin{subequations}
		\begin{align}
			\xi_s \dot{z}_s(t) & = + \pi \Gamma \left ( R^2 - \zeta(z_s,z_l)^2 \right ) \cdot \Theta \left [ R - |\zeta(z_s,z_l)|
\right ] - 4 \frac{h}{\lambda_B^4} z_s (z_s^2 - \lambda_B^2) - b + F_s(t) \label{eq5a}\\
			\xi_l \dot{z}_l(t) & = - \pi \Gamma \left ( R^2 - \zeta(z_s,z_l)^2 \right ) \cdot \Theta \left [ R - |\zeta(z_s,z_l)|
\right ] + F_l(t) \label{eq5b}
		\end{align}
	\label{eq5}
	\end{subequations}
\vspace{0.3cm} 
\end{widetext}
with $\xi_x$ as friction coefficients, $x=s,l$,  and $F_x(t)$ denoting $\delta$-correlated random forces fulfilling fluctuation-dissipation theorem 
$\langle F_{x}(t) F_{x}(t_0) \rangle = 2 \mathrm{k_B T} \xi_x \delta(t - t_0)$. Note that both $\xi_s = \xi_l = 1$ since
here diffusivities of both ligand and pseudo-particle interface are set equal.

\section{Numerical simulations}

\label{sec:numerical}

\begin{figure}[t]
\includegraphics[width=8.6cm]{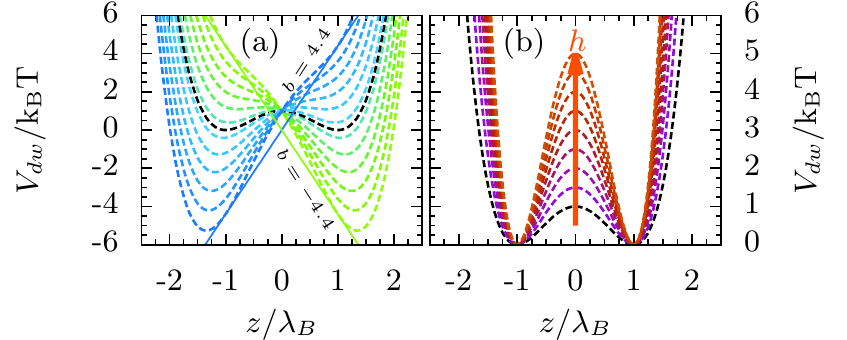}
\caption{The double-well potential which governs solvation fluctuations in the
pocket: The fluctuations are
tuned by (a) biasing from $b=-4.4~\mathrm{k_BT}/\lambda_B$ to
$4.4~\mathrm{k_BT}/\lambda_B$ in steps of
$0.8~\mathrm{k_BT}/\lambda_B$, hence breaking rate symmetry, or (b)
barrier height from $h=1~\mathrm{k_BT}$ to $5\,\mathrm{k_BT}$ in steps
of $0.5~\mathrm{k_BT}$, thus changing the magnitude of the
interface's characteristic time scales. Black dashed lines draw the reference double
well setting
($h=1$, $b=0$). Example biasing $b$ or changes in magnitude $h$ are indicated by
example guide lines (a) or arrow (b), respectively. (Color coding is
maintained throughout the paper.)}
\label{fig3}
\end{figure}

In the following we discuss binding kinetics obtained from integrating the
equations~\eqref{eq5} numerically, where we use the numeric scheme proposed by
Ermak and McCammon~\cite{Ermak&McCammon}. We focus on how their coupling affects the ligand's
reaction coordinate $z_l(t)$ kinetics, and how changes in the interface dynamics
impact. 
In general, pocket solvation can be affected by changing its hydrophobicity,
geometry or size. Such changes, however, simultaneously affect pocket occupancy and
solvent fluctuation time scale.  In our model,  we have the ability to disentangle
both effects and their influence on ligand binding.

Water occupancy can be tuned by changes in the biasing parameter $b$ of the double
well potential,  $V_{dw}$. Potentials with biasing ranging from
$b=-4.4\,\mathrm{k_BT}/\lambda_B$ to $b=4.4\,\mathrm{k_BT}/\lambda_B$ in steps of
$0.4\,\mathrm{k_BT}/\lambda_B$ are drawn in Fig.~\ref{fig3}~(a). The black dashed
line refers to the reference with barrier height $h=1\,\mathrm{k_BT}$ and no
biasing, which is motivated from explicit and implicit solvent simulation
studies~\cite{Setny:PNAS,Setny:PRL}. The barrier height can separately be tuned
from $h=1\,\mathrm{k_BT}$ to $h=5\,\mathrm{k_BT}$ in steps of $0.5\,\mathrm{k_BT}$
as plotted in Fig.~\ref{fig3}~(b). Changes in barrier height directly influence the
wet-dry transition time and thus the effective interface fluctuation time scale.
From Kramer's rate theory one knows that the rate $r$ of crossing the double well
barrier scales exponentially with barrier height $r \propto
\mathrm{e}^{-h}$~\cite{Hänggi:KramersReview}. The color coding from both plots in
Fig.~\ref{fig3} is consistently adopted to other plots throughout this paper.

\begin{figure}[t]
\centerline{\includegraphics[width=8.6cm]{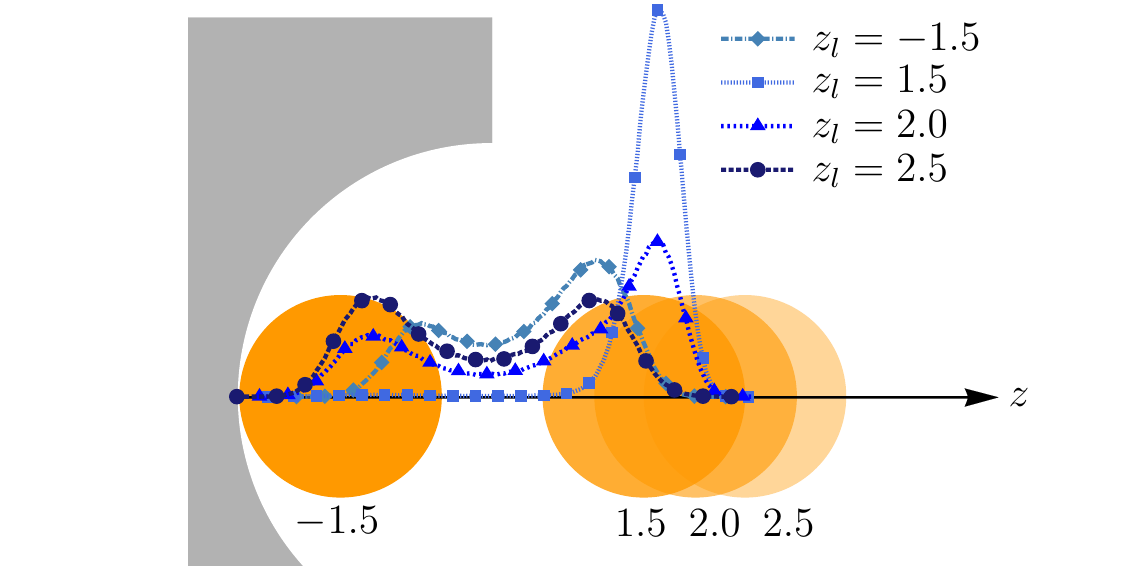}}
\caption{The equilibrium distribution (blue lines) of the water interface is affected by the ligand position due to the hydrophobic 
ligand-interface interaction in eq.~\eqref{eq3} and \eqref{eq4}. As the ligand (orange circle)
approaches the pocket, the  bimodality of the water interface distribution is lost for
intermediate states, e.g. $z_l=1.5$, but is recovered if the ligand is fully bound
to the pocket, e.g., $z_l=-1.5$.}
\label{fig4}
\end{figure}

Further we note that the equilibrium distribution of the water interface
depends on the ligand position due to the nonlinear coupling evident from
contributions of eq.~\eqref{eq4}~in~\eqref{eq5} if $|\zeta|<R$. A schematic plot in Fig.~\ref{fig4} illustrates how the bimodal
distribution changes while the ligand comes closer to the pocket. When the ligand and
the interface interact, their pair potential adds to the effective equilibrium free energy landscape 
of the pseudo-particle. This can tilt the bimodal distribution of the interface and thus influences the equilibrium 
wetting behavior of the pocket. We observe that the pocket drying is enhanced for close ligand positions 
which coarsely mimics also how pocket hydration couples to
ligand position in all-atom and implicit solvent simulations~\cite{Setny:PRL, Setny:PNAS}.

\subsection{Mean first passage time and memory}

\label{sec:MFPT}

As first measure of ligand binding kinetics the mean first passage time (MFPT)
is sampled from each point $z$ to a bound configuration at $z_f=-1.25$.
Therefor, for each setup with given biasing $b$ and barrier height $h$ around
$2\times10^{5}$ trajectories are simulated and analyzed for the ligand starting at $z_{max} = 5$ until it
is $<1.25$ inside the pocket.
Hence, the resulting MFPT curves describe the mean first passage time $T_1(z,z_f)$ of the
ligand crossing $z_f=-1.25$, given it started at $z$ with a reflective boundary at
$z_{max}=5$. 

\begin{figure}[t]
\includegraphics[width=8.6cm]{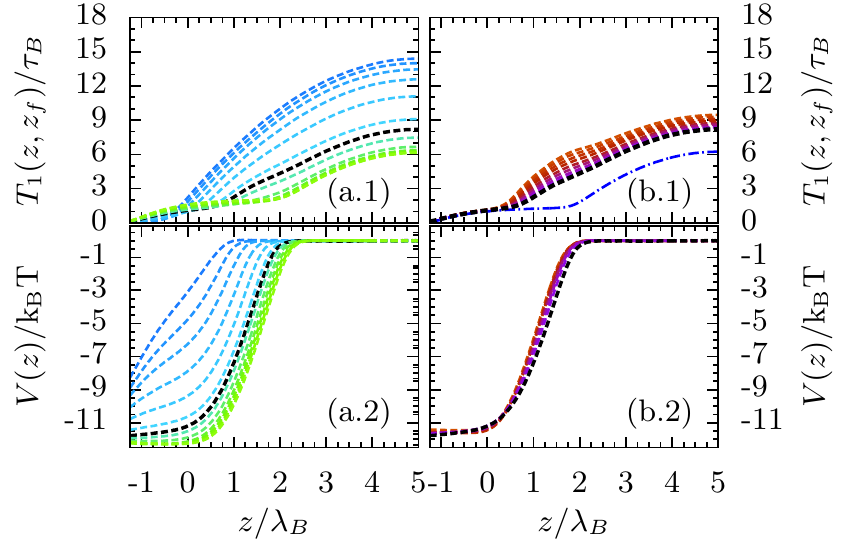}
\caption{MFPT curves $T_1(z)$ and PMF $V(z)$ of ligand binding to the pocket depend
on changes of interface's fluctuation behavior by double-well variations from
Fig.~\ref{fig3}. Panel (a.1)/(a.2) shows ligand's MFPT/PMF dependent on double-well
biasing $b$. Panel (b.1)/(b.2) draws ligand's MFPT/PMF dependent on double-well
barrier height $h$. Blue dash-dotted line in panel (b.1) draws an example MFPT
curve $T_1^\mathrm{M}(z)$ using eq.~\eqref{eq6} for the reference setting ($h=1$,
no bias) with respective PMF and constant diffusivity. (Color coding
is adopted from Fig.~\ref{fig3}.)} 
\label{fig5}
\end{figure}

Fig.~\ref{fig5}~(a.1) shows the MFPT curves corresponding to simulation setups with
varying bias in the double-well potential the interface coordinate is subject to.
Starting with setups with a negative
bias (greenish lines), hence a preferentially dry pocket, the ligand's mean binding
time is faster than without biasing (black). With growing values for the bias the MFPT slows
down by a factor of two. 
Also, as the barrier height increases, a deceleration of the MFPT is observed in
Fig.~\ref{fig5}~(b.1). There the MFPT
curves exhibit a hunch around $z\approx1$ which is enhanced with growing barrier
whereas at $z=0$ all curves $T_1(z,z_f)$ in panel (b.1) coincide.

Further, we analyze for all considered parameter values $(h,b)$ the potential of mean
force (PMF) of the ligand using the weighted histogram analysis method
(WHAM)~\cite{Ferrenberg&Swendsen,Kumar&Rosenberg,Zhu&Hummer} along $z$.
Fig.~\ref{fig5}~(a.2) shows a strong dependence of the PMFs
on the biasing parameter $b$. Besides small changes in shape, the attractive part of the PMFs
essentially shifts towards smaller values of $z$, if the interface bias shifts
towards an increasingly wetted pocket. 
On the other hand, the ligand PMF negligibly depends on the barrier height, $h$, as
revealed in
Fig.~\ref{fig5}~(b.2). The PMF, as an equilibrium quantity, is essentially
unaffected since mainly interface kinetics change with $h$. This is especially noteworthy since the
corresponding MFPT curves in panel (b.1) alter relatively strongly with $h$, suggesting
that the effect on ligand binding times originates from modified
interface kinetics. 

In the case of a Markovian process the PMF, $V(z)$, together with possibly spatially
dependent diffusivity, $D(z)$, determine the $n$-th moment of the
first passage time distribution \cite{Siegert,Weiss}
\begin{equation}
T_n^{\mathrm{M}}(z,z_f) = n \int\limits_{z_f}^{z} \mathrm{d}z' \frac{\mathrm{e}^{V(z')} }{D(z')} \int\limits_{z'}^{z_{max}}
\mathrm{d}z'' \mathrm{e}^{-V(z'')} \cdot T_{n-1}^{\mathrm{M}}.
\label{eq6}
\end{equation}
where the zeroth moment $T_{0}^{\mathrm{M}}=1$ determines normalization. Note that
$\beta = \mathrm{k_BT}^{-1} = 1$ is omitted in the Boltzmann factors in the above
equation as it will be in later occurrences. 
The blue, dash-dotted line in Fig.~\ref{fig5}~(b.1) is a numerically integrated
solution of equation~\eqref{eq6} using constant diffusivity $D(z)=1$, and spatially
dependent $V(z)$ of the reference case $b=0$ and $h=1$. It should be
compared to the black, dashed MFPT curve in the same plot. Only for negative $z$-values both
coincide. Effects that can no longer be treated by Markovian kinetics occur
around $z\approx1$, where the hunch in simulated MFPT curves qualitatively
deviates to a dent in the solution of equation~\eqref{eq6}. For even bigger values
of the reaction coordinate, shapes of the MFPT curves of both methods only conform, but
the Markovian solution represents overall faster association. Together the ligand dynamics
only can be modeled by pure Markovian
description when the ligand is inside the pocket.

\begin{figure}[t]
\includegraphics[width=8.6cm]{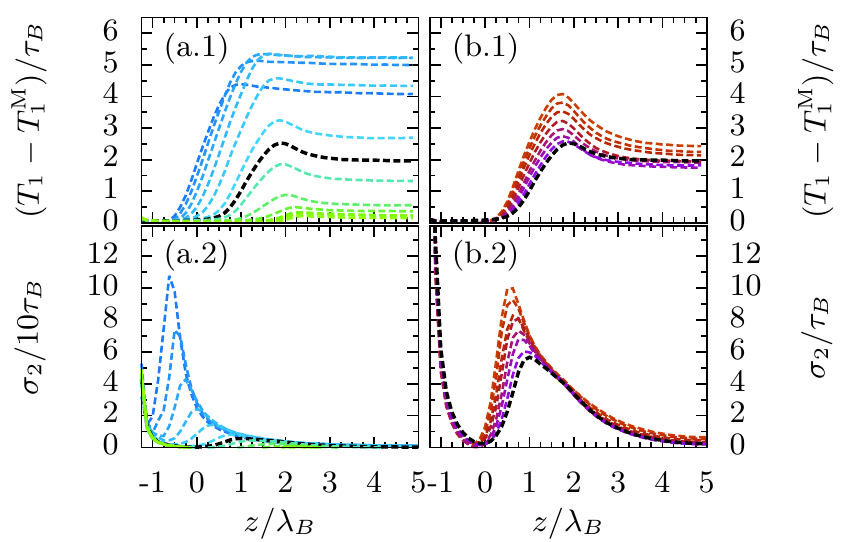}
\caption{Markovian assumption breaks down as manifested in differences of MFPT $T_1(z)$ from simulation
to $T_1^\mathrm{M}(z)$ from eq.~\eqref{eq6} and in memory index $\sigma_2(z)$ from eq.~\eqref{eq7}. Panels (a.1) and
(a.2) show the respective measures dependent on the double
well bias $b$. Panels (b.1) and (b.2) draw both measures dependent on double
well barrier height $h$. Note that the effect is much bigger as the biasing towards a
wet pocket increases. Therefore the scale of (a.2) is $\sigma_2/10\tau_B$. (Color coding is adopted from Fig.~\ref{fig3}.)}
\label{fig6}
\end{figure}

As a general measure, calculating MFPT curves $T_1^{\mathrm{M}}$ in a Markovian
picture for all
considered cases of bias strengths and barrier heights, and
using the respective PMFs, enables direct observations where the deviations in
simulation occur. Accordingly, the difference
$T_1(z,z_f)-T_1^{\mathrm{M}}(z,z_f)$ is plotted in Fig.~\ref{fig6}~(a.1) and (b.1).
In all cases the difference vanishes inside the pocket and increases
towards a maximum situated just in front of the pocket mouth. It then
plateaus to a constant positive value for large $z$, which indicates slowed ligand
kinetics in all considered cases of pocket water fluctuations. 
For the cases of biased wetting, the difference $T_1-T_1^{\mathrm{M}}$ in
Fig.~\ref{fig6}~(a.1) is very small, if the pocket is preferably
dry, namely with a strong negative bias. As the biasing parameter $b$ increases, and thus the
interface's distribution tends towards mainly hydrated pocket states, the
difference in MFPT accumulates to a peak when $b=2.8\mathrm{k_BT}/\lambda_B$, and
alleviates for even higher bias. In Fig.~\ref{fig6}~(b.1) deviations to the
Markovian picture are enhanced by growing barrier height,  hence slowed-down wetting
fluctuations.  

In order to investigate further the break-down of Markovian dynamics and possibly accompanied memory, we
additionally determine the, so called, memory index \cite{Hänggi&Talkner}
\begin{equation}
\sigma_2=\frac{T_2(z,z_f)-T_2^{\mathrm{M}}(z,z_f)}{T_1(z,z_f)}
\label{eq7}
\end{equation}
introduced by H\"anggi~\textit{et~al.}~\cite{Hänggi&Talkner} who noted its possible
value to ligand migration studies. It provides an additional spatially resolved measure of the
character of a random process. It indicates a process to be non-Markovian if the
difference of the second moment of a random process to the second moment of the
corresponding process with Markovian assumption does not vanish.
For all considered cases of pocket wetting behavior Fig.~\ref{fig6}~(a.2) and
(b.2) show that $\sigma_2$ vanish for ligand positions far from the pocket.
Note, however, that finite $\sigma_2$ values are reached already at intermediate
positions, that is $z \sim 3$, where the ligand is still out of reach of the
solvation potential ($\zeta(z_l,z_s) > R$). Finally, the memory index peaks at the
position $z\lesssim1$, at which initial deviations in the first moments occur in
Fig.\ref{fig6}~(a.1) and (b.1). 
Subsequently, for $z<1$, it steeply recedes to zero.
Inside the pocket $\sigma_2$ diverges once more, majorly due to numerical
inaccuracies close to the target distance as these are amplified by division in
eq.~\eqref{eq7}. Also, actual non-Markovian effects reoccur inside the
pocket as it becomes evident from results and discussions in following sections and
in Appendix~\ref{SI:fits}.

All together it is inept to assume constant ligand diffusivity within
a Markovian description of our ligand-interface system in order to estimate correct
mean binding time. The process
rather indicates non-Markovian contributions, slowing the ligand binding which
predominantly arise in the region where the ligand
and the interface start interacting. Overall, the first and second
moments are influenced far outside where the ligand is driven by delta correlated
random noise by definition in eq.~\eqref{eq5}. In addition, strongly dissimilar MFPTs but
almost similar PMFs in Fig.~\ref{fig5}~(b.1) and (b.2), respectively, suggest that local coupling
of interface time scales alone is enough to impact ligand binding kinetics.

\subsection{Spatially dependent friction}

\label{sec:friction}

In order to illuminate the kinetic effect we restrict further analysis to
systems considering an unbiased interface distribution, but varying only the barrier height
$h$ in the double-well potential.
For these particular cases, a purely kinetic nature of the observed effects was
evident from the perturbation of average ligand binding times occurring with
unchanged PMF, as shown in Fig. 5.b.

\begin{figure}[t]
\includegraphics[width=8.6cm]{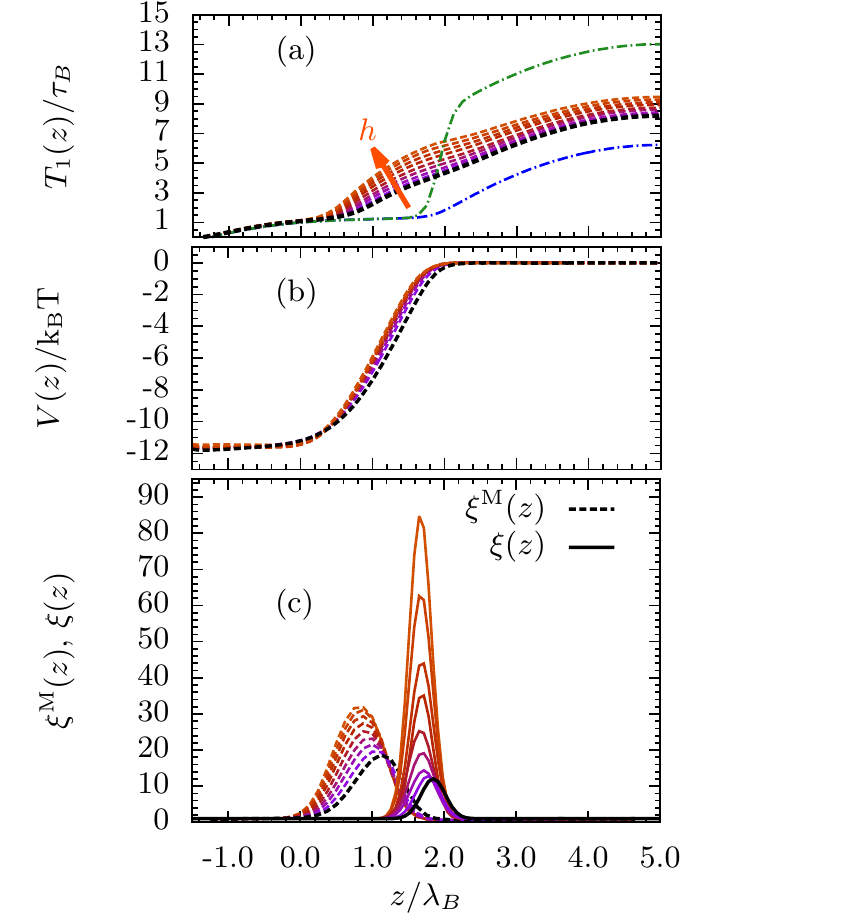}
\caption{Panel (a) draws again the MFPT curves from simulation already shown in
Fig.~\ref{fig5}~(b.1). Here it also compares to a MFPT curve $T_1^{M}(z)$ for the
reference setting calculated from eq.~\eqref{eq6} with PMF and spatially dependent
friction from PACF plotted as green dash-dotted line. Panel (b) draws again the
PMFs already shown in Fig.~\ref{fig5}~(b.2). Panel (c) plots spatially resolved
friction $\xi(z)$ from fits to PACF data eq.~\eqref{eq8} (solid) and from MFPT
data using eq.~\eqref{eq9} with Einstein relation
$\xi^\mathrm{M}(z)=k_BT/D^\mathrm{M}(z)$ (dashed). (Color coding is adopted from
Fig.~\ref{fig3}.) }
\label{fig7}
\end{figure}

Spatially resolved friction $\xi_l(z)$ can be obtained from the position
auto-correlation function (PACF), $\langle \delta z_l(t) \delta z_l(0) \rangle_r$,
with $\delta z_l(t)=z_l(t)-z_r$,  
derived from simulations with the ligand harmonically restrained at $z_r$. A detailed
description of ligand umbrella simulation is provided in Appendix~\ref{SI:fits}.
Integration of the
auto-correlation function provides the time scale which is
divided by the square of the position fluctuations, yielding the local friction \cite{Hummer}

\begin{equation}
\xi(z_r) = \frac{\int_{0}^{\infty} \langle \delta z_l(t) \delta z_l(0) \rangle_r \mathrm{d} t}{ \langle \delta z_l^2
\rangle_r^2 } \,\,\,.
\label{eq8}
\end{equation} 
Solid curves in Fig.~\ref{fig7}~(c) show Gaussian fits to
$\xi(z_r)$, gathered from PACF evaluation (see also Appendix~\ref{SI:fits}).
Simulations restraining the ligand far away from the pocket yield
the preset friction of value one. However, $\xi(z)$ strongly peaks in
front of the pocket mouth, where the ligand is subject to interaction with the
interface. Growing barrier height, and thus exponentially slowed double-well
transition rates of the interfacial motion, increases the peak up to a factor of approximately
85. It indicates that the effect arises due to the ligand
interacting with the bimodally fluctuating interface. While the interface penetrates the
pocket, the ligand, still remaining around $z \sim 2.0$, is only subject to the
$\delta$-correlated random force of the Langevin model eq.~\eqref{eq5}. Whereas when
the interface is in the outer well in front of the pocket, the 
solvation potential acts on the ligand. Hence the solvation force acts as
additional fluctuating force which introduces additional friction. Peaking friction occurs in
regions in which fluctuations are most pronounced, where interface and ligand might
interact or not, and thus the solvation potential can essentially be \textit{on} or
\textit{off}. 

Together with the observations throughout previous MFPT analysis it
seems that the additional force fluctuations serve as a source of memory. Their time scale is proportional
to the Brownian time, $\tau_B$, and scales exponentially with double well barrier
height:  $\tau_{dw}\propto\tau_B\mathrm{e}^h$ (see also eq.\eqref{eq17}). 
Thus the time scale of the solvation force fluctuations does not separate from the time scale
of ligand migration, which leaves memory to the association process. As it will be shown in the following section a
\textit{generalized Langevin model} derives an exponential growth of the friction
peak values with barrier height, which directly relates to exponentially growing
time scales of the interface fluctuations.

Again we obtain the MFPT curve $T_1^\mathrm{M}(z,z_f)$ using equation~\eqref{eq6},
with the PMF $V(z)$ of the reference case $(h=1,b=0)$, and now additionally with a
spatially resolved diffusion $D(z)~=~\mathrm{k_BT}~\cdot~\xi^{-1}(z)$ from Einstein's
relation, with previously evaluated $\xi(z)$. Only in the interval
$z\in (-1.25,0.5)$ the result, plotted as green
dash dotted line in Fig.~\ref{fig7}~(a), coincides with MFPT curves from simulation
(black dashed), and moreover, with evaluation of eq.~\eqref{eq6}
without spatially resolved friction (blue dash dotted). Subsequently, a steep edge
in the curve yields values which overestimate the actually simulated results far
outside, $z\gg2$. So, on one hand, the solution of equation~\eqref{eq6} overestimates the
results using both spatially resolved profiles $V(z)$ and $\xi(z)$. On the other
hand, it is underestimated using only spatially resolved PMF, but constant friction
of value one.

For comparison we also calculate spatially dependent profiles
$\xi^{\mathrm{M}}(z)=\mathrm{k_BT}/~D^{\mathrm{M}}(z)$ by solving
equation~\eqref{eq6} for $D(z) \equiv D^{\mathrm{M}}(z)$ as primarily introduced by
Hinczewski~\textit{et~al.}~\cite{Hinczewski&Netz}
\begin{equation}
D^{\mathrm{M}}(z)= \frac{\mathrm{e}^{V(z)}}{\partial T_1(z)/\partial z}
\int_{z}^{z_\mathrm{max}}\,\mathrm{d}z'\,\mathrm{e}^{-V(z')}.
\label{eq9}
\end{equation}
Note that $\xi^{\mathrm{M}}(z)$ uses the Markovian assumption, and thus, is certainly
not the proper friction profile fulfilling fluctuation-dissipation theorem for our
non-Markovian ligand migration process. In detail, it does
not measure the quantity friction/dissipation which can be proportionally related
to the system's fluctuations. However, it will trivially reproduce the
correct MFPT $T_1$ from simulation when using it in eq.\eqref{eq6}. Dashed lines in
Fig.~\ref{fig7}~(c) show curves for $\xi^\mathrm{M}(z)$ which similarly peak in front
of the pocket mouth. In exact comparison to $\xi(z)$ from PACF, the results
from eq.~\eqref{eq9} show a positional shift closer to the
pocket and differ in peaking value as well as peak width. (Additional information on
how $\xi^\mathrm{M}(z)$ depends on the choice of reflective boundary is discussed
in the Appendix~\ref{SI:size}.)

Again, observing essential discrepancy between local friction $\xi(z)$ from simulation
and a spatial profile $\xi^\mathrm{M}(z)$ from a memoryless picture in eq.~\eqref{eq9}
illustrates the location and strength of non-Markovian effects within the Langevin
system~\eqref{eq5}.
Also, whether calculating MFPTs from eq.~\eqref{eq6} either with constant friction
or spatially resolved friction $\xi(z)$, yields under- and overestimated results, respectively. In
neither case the Markovian property is fulfilled, which is the requirement for
obtaining a proper system description with eq.~\eqref{eq6}.
We emphasize that our system exhibits similar non-Markovian effects as those resolved by
explicit water simulations from Setny~\textit{et~al.}~\cite{Setny:PNAS}. 

\section{Generalized Langevin model}

\label{sec:GLE}

Having identified fluctuations of the solvation potential as the origin of local memory and
friction in the ligand's reaction coordinate we show in the following how their
impact can be quantified. Further, this section shall formulate the proper stochastic
characteristics when dealing with the ligand coordinate alone. For simplicity
it focuses on local conditions, namely, constrained ligand position. Yet it
elucidates a proper non-Markovian formulation to classify possible treatment with
conventional theory.

Generally in the case of a known memory kernel $\eta(t)$ one
can directly investigate the corresponding one-dimensional general Langevin equation (GLE)
\begin{equation}
m\ddot{q}(t) = -\frac{\partial V_{eq}(q)}{\partial q} - \int^t \mathrm{d}t' \eta(t-t')\dot{q}(t') + \mathcal{F}(t)
\label{eq10}
\end{equation}
with mass $m$, equilibrium potential $V_{eq}(q)$, and a random
force fulfilling fluctuation dissipation $\langle
\mathcal{F}(t)\mathcal{F}(t')\rangle=2\mathrm{k_BT}\eta(t-t')$. Simple systems
of two coupled Langevin equations can be analytically contracted onto a one-dimensional
GLE \cite{Risken,vanKampen} and \textit{vice versa}. A prominent example is that of an underdamped Brownian
particle in a harmonic potential. 
For the coupled system described by
eq.~\eqref{eq5} analytic contraction from 2D to 1D is not
feasible due to higher than harmonic coupling and nonlinearity in the double-well
potential. Therefor we reinterpret a method which is usually used to expand a
one-dimensional GLE to a set of two coupled equations without memory. With that we
are able to approximate friction from local conditions of the pocket-ligand system.
We restrict the analysis to the location of the friction peaks discussed above and
can predict the peaking value $\mathrm{max}( \xi(z) )$ as function of barrier
height in a bimodally fluctuating force. 

To this end we reverse the approach from
Pollak~\textit{et~al.}~\cite{Risken,Pollak&Berezhkovsii,Berezhkovsii&Pollak}. It
originally extends a one-dimensional GLE of reaction coordinate $q$ such as
eq~\eqref{eq10} by an
auxiliary variable $x$ to receive two coupled equations. Each of the resulting
equations then omits
memory and only the auxiliary variable $x$ is driven by a temporally delta correlated random
force, $N(t)$. Taking unit mass $m=1$, the GLE~\eqref{eq10} is mapped on the two dimensional,
underdamped system
\begin{subequations}
	\begin{align}
	\label{eq11a}
	\ddot{q} + \frac{\partial V(q,x)}{\partial q} &= 0 \\
	\label{eq11b}
	\xi \dot{x} + \frac{\partial V(q,x)}{\partial x} &= N(t).
	\end{align}
\label{eq11}
\end{subequations}
The driving noise $N(t)$ is a delta correlated Gaussian noise
\begin{equation}
\langle N(t) \rangle = 0, \,\,\,\, \langle N(t) N(t') \rangle =
2\mathrm{k_BT}\xi~\delta(t-t') \,.
\label{eq12}
\end{equation}
There are two further requirements that memory $\eta(t)$ and coupling potential $V(q,x)$ must
fulfill for proper mapping~\cite{Pollak&Berezhkovsii,Berezhkovsii&Pollak}:
\begin{enumerate}
\item[(a)]{The kernel $\eta(t)$ may be represented by a sum of exponentials, and for this very example even
	\begin{equation}
		\eta(t)=\frac{\xi}{\tau} \mathrm{e}^{-t/\tau} \equiv \Omega
\mathrm{e}^{-t/\tau} .
		\label{eq13}
	\end{equation}}
\item[(b)]{The coupling between auxiliary and reaction coordinate should be harmonic such that
	\begin{equation} 
		\frac{\partial V(q,x)}{\partial q} = \frac{\mathrm{d} V_{eq}(q)}{\mathrm{d} q} - \Omega [x -
f(q)]\frac{\mathrm{d}f(q)}{\mathrm{d}q} . 
	\label{eq14}
	\end{equation}
	\vspace{0.25cm}
}
\end{enumerate}

\begin{figure}[b]
\includegraphics[width=8.6cm]{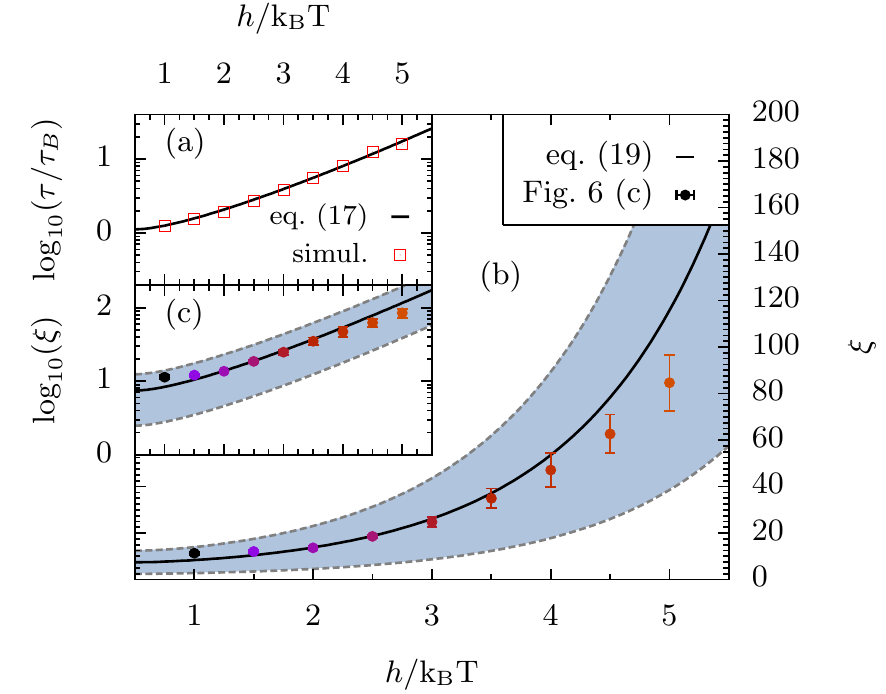}
\caption{Panel (a) plots time scale $\tau_{dw}(h)$ of the interfacial motion in the double-well
without ligand. Black line refers to eq.~\ref{eq17} and orange squares are
values obtained by interface's position auto-correlation from simulation. Panel (b) shows
$\xi(h)$~(eq.~\eqref{eq19}) from constructed GLE using $\epsilon=0.36$
(black line) and blueish shaded area delimited by gray dashed lines indicates the
range covered by eq.~\ref{eq19} with standard deviation
$\Delta\epsilon=\pm0.24$. The peaking values from
$\xi(z)$ in Fig.~\ref{fig7}~(c) are here shown as circular symbols with
corresponding color coding from Fig.~\ref{fig7}. Panel (c) plots the same data as in (b) on a log-scale. }
\label{fig8}
\end{figure}
In our case, let us focus on the situation at the position of the friction peak, $\mathrm{max}(
\xi(z) )$ in Fig.~\ref{fig7}~(c).
In that case the expansion $\mathcal{T}_1^\epsilon$ of the solvation
force in eq.~\eqref{eq4} at fixed ligand-interface distance $\epsilon$ up to first order, with respect to a perturbation
$\delta\zeta = \zeta-\epsilon$, gives the harmonic contribution of our solvation coupling
\begin{equation}
\mathcal{T}_1^\epsilon~F_{sol}(\zeta) = \pi\Gamma [\epsilon^2-R^2] + 2\pi\Gamma\epsilon (\zeta - \epsilon) + \mathcal{O}(\zeta^2)
\label{eq15}
\end{equation}
which identifies the memory kernel constant, $\Omega=\xi/\tau=2\pi\Gamma\epsilon$, by
comparison with eq.~\eqref{eq14}. The value of
$\epsilon\approx0.36\pm0.24$ is estimated from simulations constraining the ligand at the
position of the friction peaks in Fig.~\ref{fig7}. It is the mean and standard deviation of the
distribution of distance $\zeta$ between constrained ligand and bimodally
fluctuating interface. Detailed evaluations of $\epsilon$ are
discussed in Appendix~\ref{SI:peak}. The set of coupled equations of motion of a
free ligand (here $q_l$)
coupled to an auxiliary variable $x_s$ are adopted to the
requirements (a) and (b) described above such that
\begin{subequations}
	\begin{align}
	\label{eq16a}
	\ddot{q}_l - 2\pi\Gamma\epsilon(q_l - x_s - R) &= 0 \\
	\label{eq16b}
	\xi \dot{x}_s + 2\pi\Gamma\epsilon(q_l - x_s - R) &= N(t).
	\end{align}
\label{eq16}
\end{subequations} 
Note that the system of the two above equations is not equivalent to the original
coupled Langevin system \eqref{eq5}. With the aim to
formulate the influence of interface fluctuations on local friction encountered by
the ligand, it describes only a single system configuration, which determines the
$\epsilon$. 

A striking difference is that eq.~\eqref{eq16} does not implement the
double-well itself. Rather the time scale determining the memory is chosen to
be that of a Brownian particle in a double-well. A compact
approximate solution of that time scale is given by~\cite{Cregg&Wickstead, Kalmykov&Waldron}
\begin{equation}
\tau_{dw}(h) = \frac{\tau_B}{2\sqrt{2h}} \frac{(\mathrm{e}^{h} - 1)}{2h}(\pi
\sqrt{h}+ 2^{1-\sqrt{h}}) \,\,\, .
\label{eq17}
\end{equation}
To confirm the approximation for the setups we previously considered, we probe the
time scale of interface fluctuations in the double-well within
simulations without ligand. Tuning the barrier height from $h=1~\mathrm{k_BT}$ to
$h=5~\mathrm{k_BT}$ reveals that the approximate formula~\eqref{eq17} is in very
good agreement within the range of interest in $h$ (Fig.~\ref{fig8}~(a)).

The GLE corresponding to \eqref{eq16} with memory from
\eqref{eq17} is followingly given by 
\begin{equation}
\ddot{q}_l(t) = - 2\pi\Gamma\epsilon \int^t \mathrm{d}t'~
\mathrm{e}^{-(t'/\tau_{dw})}\dot{q}_l(t') + \mathcal{F}(t) \,\,\,.
\label{eq18}
\end{equation}
Comparison of its memory kernel with eq.~\eqref{eq13} determines the friction for the constructed
system such that
\begin{equation}
\xi(h) = \Omega\cdot\tau = 2\pi\Gamma\epsilon\cdot\tau_{\text{dw}}(h) \, .
\label{eq19}
\end{equation} 
Fig.~\ref{fig8}~(b) and (c) demonstrate the strong resemblance of both systems, the fully
coupled key-lock binding model~\eqref{eq5} and the non-Markovian
model~\eqref{eq18}.
The circular symbols with error bars from Gaussian fits draw the maxima of the
friction peaks $\mathrm{max}(\xi(z))$ (from Fig.~\ref{fig7}) from PACF calculations of
the original key-lock model~\eqref{eq5}.
The black line draws expression~\eqref{eq19} found for $\xi(h)$. The blue
shade indicates the error from variance calculations to $\epsilon$. 

\section{Concluding Remarks}

\label{sec:conclusion}

Our investigations presented here reveal the origin of increased friction and
additional memory in hydrophobic pocket-ligand binding as it was observed in
previous work using all-atom simulations~\cite{Setny:PNAS,Mondal:PNAS}. We employ a
simple stochastic model of two nonlinearly coupled Langevin equations each driven
with \textit{memoryless Gaussian noise}. One equation models pocket hydration in
terms of a continuously diffusing pseudo-particle as an interface, which can occupy
the pocket volume or sample the region in front of the pocket entrance. Another one
describes a ligand, freely diffusing on an effectively one-dimensional
reaction-coordinate, which is subject to solvation force when in contact with the
interface. Thus, a nonlinear coupling is an
effective interaction potential between the pseudo particle and the ligand, motivated from solvation
free energy of a microscopic hydrophobic ligand scaling with the solvated volume. The model
enables investigation on tunable interface motion, hence pocket hydration, by biasing a double well potential which
leaves the pocket in rather 'wet' or rather 'dry' states, as well as by changing its barrier
height, thus affecting the respective transition rate.
Incorporating the double-well behavior of pocket hydration was essentially
motivated from the bimodal water occupancy distributions observed in hydrophobic
confinement by Setny~\textit{et al.}~\cite{Setny:PNAS}.

Even though the system was driven by simple Markovian delta-correlated random
forces, the first passage time analysis of ligand binding revealed non-Markovian
contributions to binding kinetics. In all cases of bimodal water interface
fluctuations, ligand association kinetics from numerical simulation of the model
was decelerated in comparison to a Markovian picture, utilizing equilibrium PMF and
spatially constant friction. Deviations of numerical simulations from the Markovian
description were spatially resolved for comparison of the first and the second moments of the
first passage times of binding to the pocket. Comparison of the moments indicated
non-Markovian contributions to occur shortly before binding, at positions where the
ligand is subject to intermittent interaction with the bimodally fluctuating interface.
Otherwise the Markovian behavior is restored if the ligand
is already inside the pocket, where the pair potential inhibits bimodal
interface fluctuations and leaves the pocket in a rather dry state.
 
Especially when hydration fluctuation time scales were changed by tuning double-well
barrier height, the ligand PMFs basically remained
unchanged. At the same time, the deceleration of ligand binding was enhanced by
increasing interface relaxation times.
Together it raised evidence that ligand kinetics couples to the time scale of water
fluctuations due to the fluctuating potential, which in all-atom simulations refers to
a fluctuating PMF~\cite{Mondal:PNAS}. 

We resolved friction of the ligand by position auto-correlation and found spatially
dependent friction in front of the pocket. It was found to peak at positions
where coupling to the water interface occurs time-dependently. As the ligand is situated at
the edge of the interface's external double-well it is subject to the solvation
potential only for given time periods. The situation exhibits a bimodally
\textit{on-off} switching of coupling and thus facilitates additional force
fluctuations, yielding additional friction.

We corroborate the origin of memory by constructing a generalized Langevin model
restricted to local conditions of the original two coupled Langevin equations.
It utilized the interaction potential between the ligand and the interface, as well as the auto-correlation time
of the interface fluctuation as time scale of the memory kernel. The derived friction and noise
strength of the GLE~\eqref{eq18} was shown to coincide with the observed modulations in spatial ligand
friction from simulations of eq.~\eqref{eq5}.

In general, additional friction and memory in hydrophobic key-lock binding
can originate from coupling of water fluctuations to the ligand diffusion when both
occur on comparable time scales, as shown here and in previous MD studies
\cite{Setny:PNAS,Mondal:PNAS}. Essentially, the pocket water fluctuations behave as strong
and comparably slow fluctuations of the mean forces (PMF) to the ligand. Within the course of this
paper especially bimodal fluctuations of the coupling facilitate strong \textit{on-off} force
fluctuations. These do not rank as the fast
solvent molecular forces for which ligand and solvent time scales separate which can be
coarse-grained in a random force kernel delta-correlated in time. Instead, the slow solvent fluctuations in the pocket rather add to ligand friction including memory as fluctuation-dissipation theorem
predicts, dissipative forces (friction) to be proportionally related to a system's intrinsic
fluctuations.

With this the paper illustrated the kinetic characteristics of ligand association coupling to
pocket water
occupancy fluctuations. It suggests to future studies on ligand-receptor systems to apply
elements of conventional kinetic theory which also accounts for situations when time scales do not
separate. For extreme cases with bimodal
wetting fluctuations, a two-state approach has been successfully applied
in ref.~\cite{Mondal:PNAS}, suggesting possible consideration in reaction-diffusion
models, multistate models \cite{Zhou,Szabo:JCP} or even Markov-state models
\cite{Bowman&Noe,Hummer&Szabo}. Other studies \cite{Mondal:JCTC} make evident that
ligand binding is not necessarily accompanied by bimodal pocket occupancy
fluctuations. If, in such cases, memory remains corrections to a
Kramer's rate of ligand binding and unbinding, can be determined within Grote-Hynes
theory~\cite{Grote&Hynes} and generalizations of it \cite{Hänggi}. Beyond known kinetic approaches to
ligand-receptor binding it remains to
future work to investigate fundamental and worthwhile treatment considering position dependent
memory as a ligand's position influences the (water) fluctuations it couples to.
 
\acknowledgments

R. Gregor Wei{\ss}, Piotr Setny, and Joe Dzubiella wish to express their sincere gratitude 
to their mentor J. Andrew 'Andy' McCammon for all the exciting opportunities in science, 
his never-ending support, and his kind hospitality in San Diego, California.  
The authors thank the Deutsche Forschungsgemeinschaft (DFG)  for financial support of this project. 

\begin{appendix}

\section{Units and constants}

\label{SI:units}

The Langevin system in eq.~\eqref{eq5} is strongly inspired by the all-atom simulation setup in ref~\cite{Setny:PNAS}.
Therefore all units are directly related to the system of the corresponding MD setup. Those were conducted in ambient conditions
which is why effective temperature during the numerical simulations of eq.~\eqref{eq5} relates to $T=300$~K. The particle size
of the ligand is that of a methane molecule which is roughly $R=0.4$~nm which sets the Brownian length scale
$\lambda_B$. With viscosity $\eta\approx10^{-3}\,$Pa$\cdot$s of water \cite{Smith:SPCEvisco} the corresponding diffusion constant
relates to $D=\mathrm{k_BT}/6\pi\eta R = 0.54\,\mathrm{nm}^2\,\mathrm{ns}^{-1}$. 

For parametrization of the solvation potential we assume the crossover length-scale at $l_c=1$~nm which relates to $l_c=2.5\,\lambda_B$.
With surface tension $\gamma\approx15.36\,\mathrm{k_BT}/\mathrm{nm}^{2}$ for water
\cite{Vega:SPCEtens} we calculate the
solvation volume scaling constant $\Gamma=3\gamma/l_c = 46.1\,\mathrm{k_BT}/\mathrm{nm}^3=2.95\,\mathrm{k_BT} / \lambda_B^{3}$.

\section{Umbrella sampling simulations}

\label{SI:fits}

\begin{figure}[t]
\includegraphics[width=8.6cm]{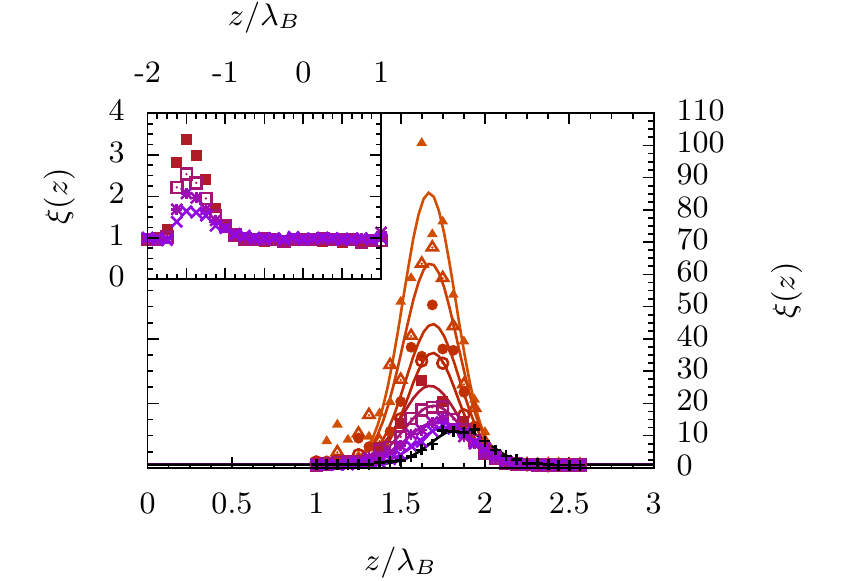}
\caption{Symbols represent spatial friction coefficients $\xi(z_r)$ at restraining positions $z_r$. Peaking values increase with
increasing double-well barrier height represented in varying color. Gaussian fits of to the
respective data are plotted as solid lines. The \textit{inset} shows similarly peaking frictions when the ligand is inside the
pocket and thus randomly interacts with the interface, whereas here $h=(1.5,~2,~2.5,~3)~\mathrm{k_BT}$. (Color coding is adopted from Fig.~\ref{fig3}.)}
\label{fig9}
\end{figure}

For a given set of parameters for eq.~\eqref{eq5} umbrella sampling simulations were
conducted. The simulations implemented harmonic
restraining forces $F_H = -k(z(t)-z_r)$ as external force to the ligand with a
spring constant of $k=53.92~\mathrm{k_BT}/\lambda_B^2$.

For PMF calculation umbrella histograms with a resolution of $0.025$ were generated from $z=-2$ to $z=5.25$ in steps of $0.125$.
In house software implementing WHAM was used to determine the unbiased equilibrium distribution from the biased umbrella
sampling distribution yielding the PMFs by Boltzmann inversion. 

For the purpose of friction calculations $\xi(z_r)$ by means of PACF umbrella
setups were restraining the ligand at positions $z_r = 1$ to $2.625$ in steps of
$0.125$. The choice of the interval was made by an initial coarse scan from
positions $z_r$ deep inside the pocket up to distances far away. The resulting
values $\xi(z_r)$ were fitted by a Gaussian illustrated in Fig.~\ref{fig9} for
simulations utilizing barrier height $h=1~\mathrm{k_BT}$ to $5~\mathrm{k_BT}$ in
steps of $0.5~\mathrm{k_BT}$. A second friction
peak was also found inside the
pocket around $z=-1.5$ as plotted in the inset of Fig.~\ref{fig9}. The
$h$-dependence is similar because the essential underlying reason is the same but
it is not of further relevance to our discussion. A doubled spring constant gave
similar $\xi(z_r)$ within errors of approximately $5\%$ thus confirming sufficient
choice of the spring constant. 

Also note that sampling has to be increased when
barrier height was increased in order to sufficiently sample slowed water 
fluctuations. Elongated simulations were performed for statistically converged PACF
calculation. Still, however, the data remains more noisy for
simulations with extended water fluctuation time scales.

\section{\texorpdfstring{System size dependence of $\xi^{\mathrm{M}}(z)$ }{System size dependence of xi\textasciicircum M (z)}}
\label{SI:size}

\begin{figure}[t]
\includegraphics[width=8.6cm]{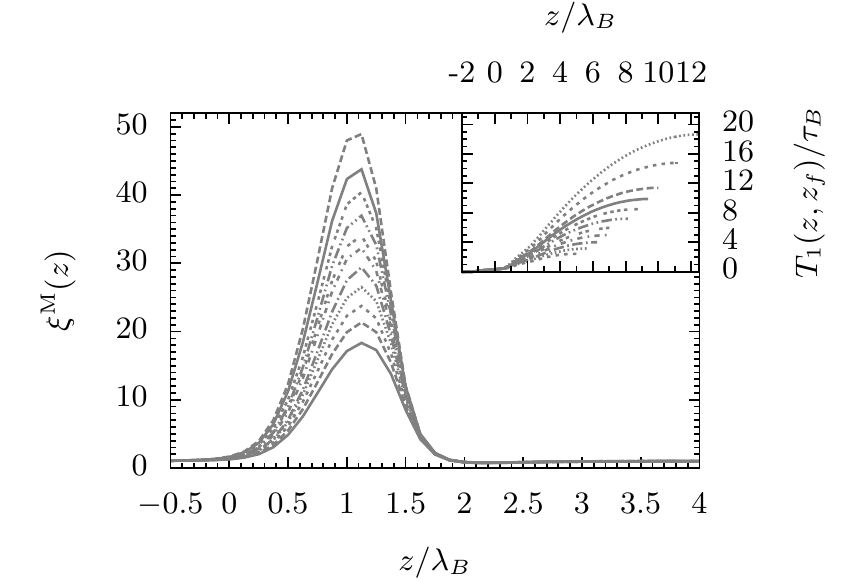}
\caption{The peak in $\xi^\mathrm{M}(z)$-profiles increase with system size namely when
the reflective boundary $z_{max}$
outside the pocket has larger values. The peaking values are proportional to the slope
in the MFPT curves, shown in the
\textit{inset}, which increases with $z_{max}$ due to extendedly available trajectories
to the random process of ligand
migration.}
\label{fig10}
\end{figure}

The peak height of $\xi^\mathrm{M}(z)$ can easily be shown to be a system size effect
from the way the method from Smoluchowski approach is met in eq.~\eqref{eq6} and
hence eq.~\eqref{eq9}. The MFPT at each point depends on the choice of
reflective boundary $z_\mathrm{max}$ because it enters as a boundary to the
integral. This becomes most evident when one considers for example a process with
constant $V(z)=0$ and constant diffusivity $D$. Equation~\eqref{eq6} then simply
yields
\begin{equation}
T_1(z,z_f)=-1/2D~(z^2-z_f^2)+z_\mathrm{max}(z-z_f) \,.
\end{equation}
Thus, the MFPT at each position $z$ increases with $z_\mathrm{max}$ contributing to
an increase in the slope of the curve that enters eq.~\eqref{eq9}.

Certainly it can also be observed from simulations of our Langevin equations~\eqref{eq5}.
Using the Markovian approach to extract the profiles $\xi^\mathrm{M}(z)$ from PMFs
and MFPT curves utilizing eq.~\eqref{eq9} is also system size dependent.
The initial expression for $T_{1}(z,z_f)$ eq.~\eqref{eq8} assumes a diffusion
process between a perfect reflective and absorbing boundary~\cite{Siegert,Weiss}.
The absorbing boundary is implemented by terminating numerical simulations when the
ligand crosses the depth of $z_f=-1.25$ inside the pocket. The reflective boundary
at $z_{max}$ in our system is a system setup dependent feature. The MFPT from each position
within the interval
$(z_f,z_{max})$ increases with $z_{max}$ due to extendedly available trajectories
during the first passage process over $z_f$. The overall MFPT curves are increased
in value and in slope as it is plotted in the inset of Fig.~\ref{fig10}. Since the
profiles $\xi^\mathrm{M}(z)$ are proportional to the derivative $\mathrm{d}
T_1(z,z_f) / \mathrm{d} z$ the peaking values increase with $z_{max}$ as
illustrated in Fig.~\ref{fig10}.

\section{Interface ligand distance at peaking friction position}
\label{SI:peak}

\begin{figure}[t]
\includegraphics[width=8.6cm]{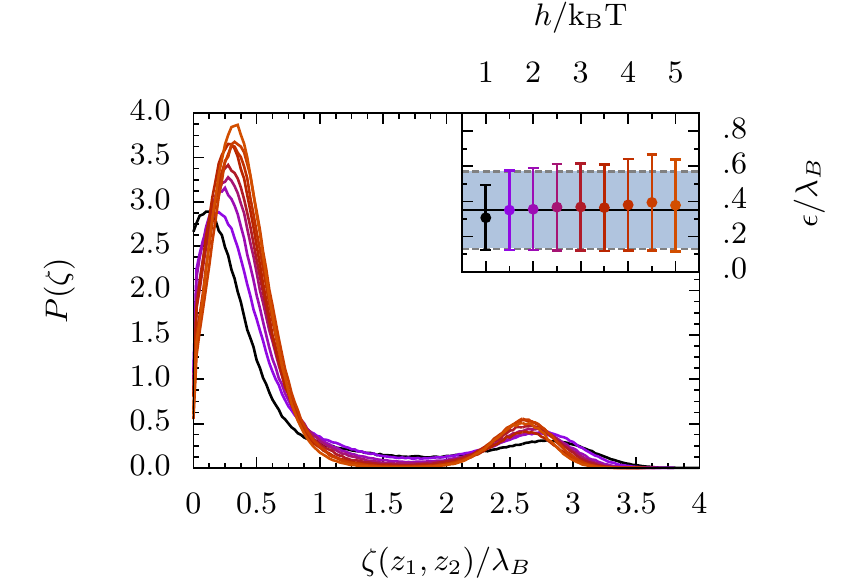}
\caption{The distributions $P(\zeta)$ are drawn for varying $h$ while the ligand is restrained at the peaking position of the
friction from Fig.~\ref{fig7}. The \textit{inset} plots the respective first moment with second moment as errorbars
evaluated from $P(\zeta)$ in the interval $[0,1]$. The average first moment is drawn as black line and the blueish shade
illustrates the average second moment. (Color coding is adopted from Fig.~\ref{fig3}.)}
\label{fig11}
\end{figure}

For development of a generalized Langevin model in the main text we expanded the
coupling due to solvation potential up to first order in $\zeta(z_1,z_2)$. It was
used to identify the force constant $\Omega = 2\pi\Gamma\epsilon \equiv \xi/\tau$
to construct a harmonic coupling between reaction coordinate $q$ and auxiliary
variable $x$ which directly relates to the noise strength of the corresponding GLE.
For comparison to the friction
values of the peaking friction from simulation the average and thus dominant
distance $\epsilon$ between interface and the ligand has to be extracted from simulation. Therefore
the ligand is fixed by an umbrella potential at the positions of each
friction peak from simulations with increasing barrier height $h$ and the
distribution $P_h(\zeta)$ is sampled. The distributions dependent on $h$ are
plotted in Fig.~\ref{fig10} whereas all $P_h(\zeta)$ behave equivalently. The
average and the standard deviation
\begin{align}
\epsilon(h) &= \langle \zeta \rangle \\
\Delta\epsilon(h) &= \langle \zeta^2 - \langle \zeta \rangle^2 \rangle^{1/2}
\end{align}
are calculated within the interval $[0,1]$ in which interface and ligand interact.
The values $\epsilon(h)$ are plotted in the inset of Fig.~\ref{fig10}
with their standard deviation as errorbars. The black line draws the average
$\epsilon = \langle \epsilon(h) \rangle_h = 0.36$ which is used in the coupling
strength. Similarly the average standard deviation $\delta\epsilon = \langle \Delta
\epsilon(h) \rangle_h = 0.24$ is plotted as blueish shade in Fig.~\ref{fig10}.

\end{appendix}

\bibliography{references}

\begin{thebibliography}{10}%
\makeatletter
\providecommand \@ifxundefined [1]{%
 \ifx #1\undefined \expandafter \@firstoftwo
 \else \expandafter \@secondoftwo
\fi
}%
\providecommand \@ifnum [1]{%
 \ifnum #1\expandafter \@firstoftwo
 \else \expandafter \@secondoftwo
\fi
}%
\providecommand \enquote [1]{``#1''}%
\providecommand \bibnamefont  [1]{#1}%
\providecommand \bibfnamefont [1]{#1}%
\providecommand \citenamefont [1]{#1}%
\providecommand\href[0]{\@sanitize\@href}%
\providecommand\@href[1]{\endgroup\@@startlink{#1}\endgroup\@@href}%
\providecommand\@@href[1]{#1\@@endlink}%
\providecommand \@sanitize [0]{\begingroup\catcode`\&12\catcode`\#12\relax}%
\@ifxundefined \pdfoutput {\@firstoftwo}{%
 \@ifnum{\z@=\pdfoutput}{\@firstoftwo}{\@secondoftwo}%
}{%
 \providecommand\@@startlink[1]{\leavevmode\special{html:<a href="#1">}}%
 \providecommand\@@endlink[0]{\special{html:</a>}}%
}{%
 \providecommand\@@startlink[1]{%
  \leavevmode
  \pdfstartlink
   attr{/Border[0 0 1 ]/H/I/C[0 1 1]}%
   user{/Subtype/Link/A<</Type/Action/S/URI/URI(#1)>>}%
  \relax
 }%
 \providecommand\@@endlink[0]{\pdfendlink}%
}%
\providecommand \url  [0]{\begingroup\@sanitize \@url }%
\providecommand \@url [1]{\endgroup\@href {#1}{\urlprefix}}%
\providecommand \urlprefix [0]{URL }%
\providecommand \Eprint[0]{\href }%
\@ifxundefined \urlstyle {%
  \providecommand \doi [1]{doi:\discretionary{}{}{}#1}%
}{%
  \providecommand \doi [0]{doi:\discretionary{}{}{}\begingroup
  \urlstyle{rm}\Url }%
}%
\providecommand \doibase [0]{http://dx.doi.org/}%
\providecommand \Doi[1]{\href{\doibase#1}}%
\providecommand \bibAnnote [3]{%
  \BibitemShut{#1}%
  \begin{quotation}\noindent
    \textsc{Key:}\ #2\\\textsc{Annotation:}\ #3%
  \end{quotation}%
}%
\providecommand \bibAnnoteFile [2]{%
  \IfFileExists{#2}{\bibAnnote {#1} {#2} {\input{#2}}}{}%
}%
\providecommand \typeout [0]{\immediate \write \m@ne }%
\providecommand \selectlanguage [0]{\@gobble}%
\providecommand \bibinfo [0]{\@secondoftwo}%
\providecommand \bibfield [0]{\@secondoftwo}%
\providecommand \translation [1]{[#1]}%
\providecommand \BibitemOpen[0]{}%
\providecommand \bibitemStop [0]{}%
\providecommand \bibitemNoStop [0]{.\EOS\space}%
\providecommand \EOS [0]{\spacefactor3000\relax}%
\providecommand \BibitemShut [1]{\csname bibitem#1\endcsname}%
\bibitem{Setny:PNAS}%
  \BibitemOpen
  \bibfield{author}{%
  \bibinfo {author} {\bibfnamefont{P.}~\bibnamefont{Setny}}, \bibinfo {author}
  {\bibfnamefont{R.}~\bibnamefont{Baron}}, \bibinfo {author}
  {\bibfnamefont{P.}~\bibnamefont{Kekenes-Huskey}}, \bibinfo {author}
  {\bibfnamefont{J.~A.}\ \bibnamefont{McCammon}},\ and\ \bibinfo {author}
  {\bibfnamefont{J.}~\bibnamefont{Dzubiella}},\ }%
  \bibfield{journal}{%
  \bibinfo {journal} {Proc. Natl. Acad. Sci. (USA)}\ }%
  \textbf{\bibinfo {volume} {110}},\ \bibinfo {pages} {1197} (\bibinfo {year}
  {2013})%
  \bibAnnoteFile{NoStop}{Setny:PNAS}%
\bibitem{Mondal:PNAS}%
  \BibitemOpen
  \bibfield{author}{%
  \bibinfo {author} {\bibfnamefont{J.}~\bibnamefont{Mondal}}, \bibinfo {author}
  {\bibfnamefont{J.~A.}\ \bibnamefont{Morrone}},\ and\ \bibinfo {author}
  {\bibfnamefont{B.~J.}\ \bibnamefont{Berne}},\ }%
  \bibfield{journal}{%
  \bibinfo {journal} {Proc.~Natl.~Acad.~Sci.~(USA)}\ }%
  \textbf{\bibinfo {volume} {110}},\ \bibinfo {pages} {33} (\bibinfo {year}
  {2013})%
  \bibAnnoteFile{NoStop}{Mondal:PNAS}%
\bibitem{Gellman}%
  \BibitemOpen
  \bibfield{author}{%
  \bibinfo {author} {\bibfnamefont{S.~H.}\ \bibnamefont{Gellman}},\ }%
  \bibfield{journal}{%
  \bibinfo {journal} {Chem.~Rev.}\ }%
  \textbf{\bibinfo {volume} {97}},\ \bibinfo {pages} {1231} (\bibinfo {year}
  {1997})%
  \bibAnnoteFile{NoStop}{Gellman}%
\bibitem{Gilson&Zhou}%
  \BibitemOpen
  \bibfield{author}{%
  \bibinfo {author} {\bibfnamefont{M.~K.}\ \bibnamefont{Gilson}}\ and\ \bibinfo
  {author} {\bibfnamefont{H.-X.}\ \bibnamefont{Zhou}},\ }%
  \bibfield{journal}{%
  \bibinfo {journal} {Annu.~Rev.~Biophys.~Biomol.~Struct}\ }%
  \textbf{\bibinfo {volume} {36}},\ \bibinfo {pages} {21} (\bibinfo {year}
  {2007})%
  \bibAnnoteFile{NoStop}{Gilson&Zhou}%
\bibitem{Woo&Roux}%
  \BibitemOpen
  \bibfield{author}{%
  \bibinfo {author} {\bibfnamefont{H.-J.}\ \bibnamefont{Woo}}\ and\ \bibinfo
  {author} {\bibfnamefont{B.}~\bibnamefont{Roux}},\ }%
  \bibfield{journal}{%
  \bibinfo {journal} {Proc.~Natl.~Acad.~Sci.~(USA)}\ }%
  \textbf{\bibinfo {volume} {102}},\ \bibinfo {pages} {6825} (\bibinfo {year}
  {2005})%
  \bibAnnoteFile{NoStop}{Woo&Roux}%
\bibitem{Wang&Wang}%
  \BibitemOpen
  \bibfield{author}{%
  \bibinfo {author} {\bibfnamefont{J.}~\bibnamefont{Wang}}, \bibinfo {author}
  {\bibfnamefont{X.}~\bibnamefont{Zheng}}, \bibinfo {author}
  {\bibfnamefont{Y.}~\bibnamefont{Yang}}, \bibinfo {author}
  {\bibfnamefont{D.}~\bibnamefont{Drueckhammer}}, \bibinfo {author}
  {\bibfnamefont{W.}~\bibnamefont{Yang}}, \bibinfo {author}
  {\bibfnamefont{G.}~\bibnamefont{Verkhivker}},\ and\ \bibinfo {author}
  {\bibfnamefont{E.~K.}\ \bibnamefont{Wang}},\ }%
  \bibfield{journal}{%
  \bibinfo {journal} {Phys.~Rev.~Lett.}\ }%
  \textbf{\bibinfo {volume} {99}},\ \bibinfo {pages} {198101} (\bibinfo {year}
  {2007})%
  \bibAnnoteFile{NoStop}{Wang&Wang}%
\bibitem{Head&Marshall}%
  \BibitemOpen
  \bibfield{author}{%
  \bibinfo {author} {\bibfnamefont{R.~D.}\ \bibnamefont{Head}}, \bibinfo
  {author} {\bibfnamefont{M.~L.}\ \bibnamefont{Smythe}}, \bibinfo {author}
  {\bibfnamefont{T.~I.}\ \bibnamefont{Oprea}}, \bibinfo {author}
  {\bibfnamefont{C.~L.}\ \bibnamefont{Waller}}, \bibinfo {author}
  {\bibfnamefont{S.~M.}\ \bibnamefont{Green}},\ and\ \bibinfo {author}
  {\bibfnamefont{G.~R.}\ \bibnamefont{Marshall}},\ }%
  \bibfield{journal}{%
  \bibinfo {journal} {J.~Am.~Chem.~Soc.}\ }%
  \textbf{\bibinfo {volume} {118}},\ \bibinfo {pages} {3959} (\bibinfo {year}
  {1996})%
  \bibAnnoteFile{NoStop}{Head&Marshall}%
\bibitem{McCammon}%
  \BibitemOpen
  \bibfield{author}{%
  \bibinfo {author} {\bibfnamefont{J.~A.}\ \bibnamefont{McCammon}},\ }%
  \bibfield{journal}{%
  \bibinfo {journal} {Curr.~Opin.~Struct.~Biol.}\ }%
  \textbf{\bibinfo {volume} {8}},\ \bibinfo {pages} {245} (\bibinfo {year}
  {1998})%
  \bibAnnoteFile{NoStop}{McCammon}%
\bibitem{Pan&Shaw}%
  \BibitemOpen
  \bibfield{author}{%
  \bibinfo {author} {\bibfnamefont{A.~C.}\ \bibnamefont{Pan}}, \bibinfo
  {author} {\bibfnamefont{D.~W.}\ \bibnamefont{Borhani}}, \bibinfo {author}
  {\bibfnamefont{R.~O.}\ \bibnamefont{Dror}},\ and\ \bibinfo {author}
  {\bibfnamefont{D.~E.}\ \bibnamefont{Shaw}},\ }%
  \bibfield{journal}{%
  \bibinfo {journal} {Drug~Discov.~Today}\ }%
  \textbf{\bibinfo {volume} {18}},\ \bibinfo {pages} {667} (\bibinfo {year}
  {2013})%
  \bibAnnoteFile{NoStop}{Pan&Shaw}%
\bibitem{Tiwary&Parrinello}%
  \BibitemOpen
  \bibfield{author}{%
  \bibinfo {author} {\bibfnamefont{P.}~\bibnamefont{Tiwary}}, \bibinfo {author}
  {\bibfnamefont{V.}~\bibnamefont{Limongello}}, \bibinfo {author}
  {\bibfnamefont{M.}~\bibnamefont{Salvalaglio}},\ and\ \bibinfo {author}
  {\bibfnamefont{M.}~\bibnamefont{Parrinello}},\ }%
  \bibfield{journal}{%
  \bibinfo {journal} {Proc.~Natl.~Acad.~Sci.}\ }%
  \textbf{\bibinfo {volume} {112}},\ \bibinfo {pages} {5} (\bibinfo {year}
  {2014})%
  \bibAnnoteFile{NoStop}{Tiwary&Parrinello}%
\bibitem{Beece&Yue}%
  \BibitemOpen
  \bibfield{author}{%
  \bibinfo {author} {\bibfnamefont{D.}~\bibnamefont{Beece}}, \bibinfo {author}
  {\bibfnamefont{L.}~\bibnamefont{Eisenstein}}, \bibinfo {author}
  {\bibfnamefont{H.}~\bibnamefont{Frauenfelder}}, \bibinfo {author}
  {\bibnamefont{D.Good}}, \bibinfo {author} {\bibfnamefont{M.~C.}\
  \bibnamefont{Marden}}, \bibinfo {author} {\bibnamefont{L.Reinisch}}, \bibinfo
  {author} {\bibfnamefont{A.}~\bibnamefont{Reynolds}}, \bibinfo {author}
  {\bibfnamefont{L.}~\bibnamefont{Sorensen}},\ and\ \bibinfo {author}
  {\bibfnamefont{K.}~\bibnamefont{Yue}},\ }%
  \bibfield{journal}{%
  \bibinfo {journal} {Biochemistry}\ }%
  \textbf{\bibinfo {volume} {19}},\ \bibinfo {pages} {23} (\bibinfo {year}
  {1980})%
  \bibAnnoteFile{NoStop}{Beece&Yue}%
\bibitem{Grote&Hynes}%
  \BibitemOpen
  \bibfield{author}{%
  \bibinfo {author} {\bibfnamefont{R.~F.}\ \bibnamefont{Grote}}\ and\ \bibinfo
  {author} {\bibfnamefont{J.~T.}\ \bibnamefont{Hynes}},\ }%
  \bibfield{journal}{%
  \bibinfo {journal} {J.~Chem.~Phys.}\ }%
  \textbf{\bibinfo {volume} {73}},\ \bibinfo {pages} {2715} (\bibinfo {year}
  {1980})%
  \bibAnnoteFile{NoStop}{Grote&Hynes}%
\bibitem{Hänggi}%
  \BibitemOpen
  \bibfield{author}{%
  \bibinfo {author} {\bibfnamefont{P.}~\bibnamefont{H{\"a}nggi}},\ }%
  \bibfield{journal}{%
  \bibinfo {journal} {J.~Stat.~Phys.}\ }%
  \textbf{\bibinfo {volume} {30}},\ \bibinfo {pages} {401} (\bibinfo {year}
  {1983})%
  \bibAnnoteFile{NoStop}{Hänggi}%
\bibitem{Doster}%
  \BibitemOpen
  \bibfield{author}{%
  \bibinfo {author} {\bibfnamefont{W.}~\bibnamefont{Doster}},\ }%
  \bibfield{journal}{%
  \bibinfo {journal} {Biophys.~Chem.}\ }%
  \textbf{\bibinfo {volume} {17}},\ \bibinfo {pages} {97} (\bibinfo {year}
  {1983})%
  \bibAnnoteFile{NoStop}{Doster}%
\bibitem{Frauenfelder&Wolynes}%
  \BibitemOpen
  \bibfield{author}{%
  \bibinfo {author} {\bibfnamefont{H.}~\bibnamefont{Frauenfelder}}\ and\
  \bibinfo {author} {\bibfnamefont{P.~G.}\ \bibnamefont{Wolynes}},\ }%
  \bibfield{journal}{%
  \bibinfo {journal} {Science}\ }%
  \textbf{\bibinfo {volume} {229}},\ \bibinfo {pages} {4711} (\bibinfo {year}
  {1985})%
  \bibAnnoteFile{NoStop}{Frauenfelder&Wolynes}%
\bibitem{Batista&Robert}%
  \BibitemOpen
  \bibfield{author}{%
  \bibinfo {author} {\bibfnamefont{P.~R.}\ \bibnamefont{Batista.}}, \bibinfo
  {author} {\bibfnamefont{G.}~\bibnamefont{Pandey}}, \bibinfo {author}
  {\bibfnamefont{P.~G.}\ \bibnamefont{Pascutti}}, \bibinfo {author}
  {\bibfnamefont{P.~M.}\ \bibnamefont{Bisch}}, \bibinfo {author}
  {\bibfnamefont{D.}~\bibnamefont{Perahia}},\ and\ \bibinfo {author}
  {\bibfnamefont{C.~H.}\ \bibnamefont{Robert}},\ }%
  \bibfield{journal}{%
  \bibinfo {journal} {J.~Chem.~Theory ~Comput.}\ }%
  \textbf{\bibinfo {volume} {7}},\ \bibinfo {pages} {2348} (\bibinfo {year}
  {2011})%
  \bibAnnoteFile{NoStop}{Batista&Robert}%
\bibitem{Ishikawa&Fayer}%
  \BibitemOpen
  \bibfield{author}{%
  \bibinfo {author} {\bibfnamefont{H.}~\bibnamefont{Ishikawa}}, \bibinfo
  {author} {\bibfnamefont{K.}~\bibnamefont{Kwak}}, \bibinfo {author}
  {\bibfnamefont{J.~K.}\ \bibnamefont{Chung}}, \bibinfo {author}
  {\bibfnamefont{S.}~\bibnamefont{Kim}},\ and\ \bibinfo {author}
  {\bibfnamefont{M.~D.}\ \bibnamefont{Fayer}},\ }%
  \bibfield{journal}{%
  \bibinfo {journal} {Proc.~Natl.~Acad.~Sci~(USA)}\ }%
  \textbf{\bibinfo {volume} {105}},\ \bibinfo {pages} {25} (\bibinfo {year}
  {2008})%
  \bibAnnoteFile{NoStop}{Ishikawa&Fayer}%
\bibitem{Bernardi&Schulten}%
  \BibitemOpen
  \bibfield{author}{%
  \bibinfo {author} {\bibfnamefont{R.~C.}\ \bibnamefont{Bernardi}}, \bibinfo
  {author} {\bibfnamefont{I.}~\bibnamefont{Cann}},\ and\ \bibinfo {author}
  {\bibfnamefont{K.}~\bibnamefont{Schulten}},\ }%
  \bibfield{journal}{%
  \bibinfo {journal} {Biotech. for Biofuels}\ }%
  \textbf{\bibinfo {volume} {7}},\ \bibinfo {pages} {83} (\bibinfo {year}
  {2014})%
  \bibAnnoteFile{NoStop}{Bernardi&Schulten}%
\bibitem{Zhou}%
  \BibitemOpen
  \bibfield{author}{%
  \bibinfo {author} {\bibfnamefont{H.-X.}\ \bibnamefont{Zhou}},\ }%
  \bibfield{journal}{%
  \bibinfo {journal} {Biophys.~J.}\ }%
  \textbf{\bibinfo {volume} {98}},\ \bibinfo {pages} {L15} (\bibinfo {year}
  {2010})%
  \bibAnnoteFile{NoStop}{Zhou}%
\bibitem{Cai&Zhou}%
  \BibitemOpen
  \bibfield{author}{%
  \bibinfo {author} {\bibfnamefont{L.}~\bibnamefont{Cai}}\ and\ \bibinfo
  {author} {\bibfnamefont{H.-X.}\ \bibnamefont{Zhou}},\ }%
  \bibfield{journal}{%
  \bibinfo {journal} {J.~Chem.~Phys.}\ }%
  \textbf{\bibinfo {volume} {134}},\ \bibinfo {pages} {105101} (\bibinfo {year}
  {2011})%
  \bibAnnoteFile{NoStop}{Cai&Zhou}%
\bibitem{Setny:PRL}%
  \BibitemOpen
  \bibfield{author}{%
  \bibinfo {author} {\bibfnamefont{P.}~\bibnamefont{Setny}}, \bibinfo {author}
  {\bibfnamefont{Z.}~\bibnamefont{Wang}}, \bibinfo {author}
  {\bibfnamefont{L.-T.}\ \bibnamefont{Cheng}}, \bibinfo {author}
  {\bibfnamefont{B.}~\bibnamefont{Li}}, \bibinfo {author}
  {\bibfnamefont{J.~A.}\ \bibnamefont{McCammon}},\ and\ \bibinfo {author}
  {\bibfnamefont{J.}~\bibnamefont{Dzubiella}},\ }%
  \bibfield{journal}{%
  \bibinfo {journal} {Phys. Rev. Lett.}\ }%
  \textbf{\bibinfo {volume} {103}},\ \bibinfo {pages} {187801} (\bibinfo {year}
  {2009})%
  \bibAnnoteFile{NoStop}{Setny:PRL}%
\bibitem{Chandler:Nature}%
  \BibitemOpen
  \bibfield{author}{%
  \bibinfo {author} {\bibfnamefont{D.}~\bibnamefont{Chandler}},\ }%
  \bibfield{journal}{%
  \bibinfo {journal} {Nature}\ }%
  \textbf{\bibinfo {volume} {437}},\ \bibinfo {pages} {640} (\bibinfo {year}
  {2005})%
  \bibAnnoteFile{NoStop}{Chandler:Nature}%
\bibitem{Setny:JCTC:2010}%
  \BibitemOpen
  \bibfield{author}{%
  \bibinfo {author} {\bibfnamefont{P.}~\bibnamefont{Setny}}, \bibinfo {author}
  {\bibfnamefont{R.}~\bibnamefont{Baron}},\ and\ \bibinfo {author}
  {\bibfnamefont{J.~A.}\ \bibnamefont{McCammon}},\ }%
  \bibfield{journal}{%
  \bibinfo {journal} {J. Chem. Theory Comput.}\ }%
  \textbf{\bibinfo {volume} {6}},\ \bibinfo {pages} {2866} (\bibinfo {year}
  {2010})%
  \bibAnnoteFile{NoStop}{Setny:JCTC:2010}%
\bibitem{Ermak&McCammon}%
  \BibitemOpen
  \bibinfo {author} {\bibfnamefont{D.~L.}\ \bibnamefont{Ermak}}\ and\ \bibinfo
  {author} {\bibfnamefont{J.~A.}\ \bibnamefont{McCammon}}%
  \bibAnnoteFile{NoStop}{Ermak&McCammon}%
\bibitem{Hänggi:KramersReview}%
  \BibitemOpen
\bibfield{author}{%
    }%
  \bibfield{author}{%
  \bibinfo {author} {\bibfnamefont{P.}~\bibnamefont{H{\"a}nggi}}, \bibinfo
  {author} {\bibfnamefont{P.}~\bibnamefont{Talkner}},\ and\ \bibinfo {author}
  {\bibfnamefont{M.}~\bibnamefont{Borkovec}},\ }%
  \bibfield{journal}{%
  \bibinfo {journal} {Rev.~Mod.~Phys.}\ }%
  \textbf{\bibinfo {volume} {62}},\ \bibinfo {pages} {251} (\bibinfo {year}
  {1990})%
  \bibAnnoteFile{NoStop}{Hänggi:KramersReview}%
\bibitem{Ferrenberg&Swendsen}%
  \BibitemOpen
  \bibfield{author}{%
  \bibinfo {author} {\bibfnamefont{A.~M.}\ \bibnamefont{Ferrenberg}}\ and\
  \bibinfo {author} {\bibfnamefont{R.~H.}\ \bibnamefont{Swendsen}},\ }%
  \bibfield{journal}{%
  \bibinfo {journal} {Phys.~Rev.~Lett.}\ }%
  \textbf{\bibinfo {volume} {63}},\ \bibinfo {pages} {1195} (\bibinfo {year}
  {1989})%
  \bibAnnoteFile{NoStop}{Ferrenberg&Swendsen}%
\bibitem{Kumar&Rosenberg}%
  \BibitemOpen
  \bibfield{author}{%
  \bibinfo {author} {\bibfnamefont{S.}~\bibnamefont{Kumar}}, \bibinfo {author}
  {\bibfnamefont{J.~M.}\ \bibnamefont{Rosenberg}}, \bibinfo {author}
  {\bibfnamefont{D.}~\bibnamefont{Bouzida}}, \bibinfo {author}
  {\bibfnamefont{R.~H.}\ \bibnamefont{Swendsen}},\ and\ \bibinfo {author}
  {\bibfnamefont{P.~A.}\ \bibnamefont{Kollman}},\ }%
  \bibfield{journal}{%
  \bibinfo {journal} {J.~Comput.~Chem.}\ }%
  \textbf{\bibinfo {volume} {13}},\ \bibinfo {pages} {1011} (\bibinfo {year}
  {1992})%
  \bibAnnoteFile{NoStop}{Kumar&Rosenberg}%
\bibitem{Zhu&Hummer}%
  \BibitemOpen
  \bibfield{author}{%
  \bibinfo {author} {\bibfnamefont{F.}~\bibnamefont{Zhu}}\ and\ \bibinfo
  {author} {\bibfnamefont{G.}~\bibnamefont{Hummer}},\ }%
  \bibfield{journal}{%
  \bibinfo {journal} {J.~Comput.~Chem.}\ }%
  \textbf{\bibinfo {volume} {33}},\ \bibinfo {pages} {453} (\bibinfo {year}
  {2012})%
  \bibAnnoteFile{NoStop}{Zhu&Hummer}%
\bibitem{Siegert}%
  \BibitemOpen
  \bibfield{author}{%
  \bibinfo {author} {\bibfnamefont{A.~J.~F.}\ \bibnamefont{Siegert}},\ }%
  \bibfield{journal}{%
  \bibinfo {journal} {Phys.~Rev.}\ }%
  \textbf{\bibinfo {volume} {81}},\ \bibinfo {pages} {4} (\bibinfo {year}
  {1951})%
  \bibAnnoteFile{NoStop}{Siegert}%
\bibitem{Weiss}%
  \BibitemOpen
  \bibfield{author}{%
  \bibinfo {author} {\bibfnamefont{G.~H.}\ \bibnamefont{Weiss}},\ }%
  \bibfield{journal}{%
  \bibinfo {journal} {Adv. Chem. Phys.}\ }%
  \textbf{\bibinfo {volume} {13}},\ \bibinfo {pages} {1} (\bibinfo {year}
  {1966})%
  \bibAnnoteFile{NoStop}{Weiss}%
\bibitem{Hänggi&Talkner}%
  \BibitemOpen
  \bibfield{author}{%
  \bibinfo {author} {\bibfnamefont{P.}~\bibnamefont{H{\"a}nggi}}\ and\ \bibinfo
  {author} {\bibfnamefont{P.}~\bibnamefont{Talkner}},\ }%
  \bibfield{journal}{%
  \bibinfo {journal} {Phys.~Rev.~Lett.}\ }%
  \textbf{\bibinfo {volume} {51}},\ \bibinfo {pages} {25} (\bibinfo {year}
  {1983})%
  \bibAnnoteFile{NoStop}{Hänggi&Talkner}%
\bibitem{Hummer}%
  \BibitemOpen
  \bibfield{author}{%
  \bibinfo {author} {\bibfnamefont{G.}~\bibnamefont{Hummer}},\ }%
  \bibfield{journal}{%
  \bibinfo {journal} {New Journal of Physics}\ }%
  \textbf{\bibinfo {volume} {7}},\ \bibinfo {pages} {34} (\bibinfo {year}
  {2005})%
  \bibAnnoteFile{NoStop}{Hummer}%
\bibitem{Hinczewski&Netz}%
  \BibitemOpen
  \bibfield{author}{%
  \bibinfo {author} {\bibfnamefont{M.}~\bibnamefont{Hinczewski}}, \bibinfo
  {author} {\bibfnamefont{Y.}~\bibnamefont{von Hansen}}, \bibinfo {author}
  {\bibfnamefont{J.}~\bibnamefont{Dzubiella}},\ and\ \bibinfo {author}
  {\bibfnamefont{R.~R.}\ \bibnamefont{Netz}},\ }%
  \bibfield{journal}{%
  \bibinfo {journal} {J. Chem. Phys.}\ }%
  \textbf{\bibinfo {volume} {132}},\ \bibinfo {pages} {245103} (\bibinfo {year}
  {2010})%
  \bibAnnoteFile{NoStop}{Hinczewski&Netz}%
\bibitem{Risken}%
  \BibitemOpen
  \bibfield{author}{%
  \bibinfo {author} {\bibfnamefont{H.}~\bibnamefont{Risken}},\ }%
  \bibfield{journal}{%
  \bibinfo {journal} {\textit{The Fokker-Planck equation: Methods of solution
  and application}, Springer, Berlin}}%
   (\bibinfo {year} {1996})%
  \bibAnnoteFile{NoStop}{Risken}%
\bibitem{vanKampen}%
  \BibitemOpen
  \bibfield{author}{%
  \bibinfo {author} {\bibfnamefont{N.~G.}\ \bibnamefont{van Kampen}},\ }%
  \bibfield{journal}{%
  \bibinfo {journal} {\textit{Stochastic processes in physics and chemistry},
  North Holland, Amsterdam}}%
   (\bibinfo {year} {2007})%
  \bibAnnoteFile{NoStop}{vanKampen}%
\bibitem{Pollak&Berezhkovsii}%
  \BibitemOpen
  \bibfield{author}{%
  \bibinfo {author} {\bibfnamefont{E.}~\bibnamefont{Pollak}}\ and\ \bibinfo
  {author} {\bibfnamefont{A.}~\bibnamefont{Berezhkovsii}},\ }%
  \bibfield{journal}{%
  \bibinfo {journal} {J. Chem. Phys.}\ }%
  \textbf{\bibinfo {volume} {99 (2)}},\ \bibinfo {pages} {1344} (\bibinfo
  {year} {1993})%
  \bibAnnoteFile{NoStop}{Pollak&Berezhkovsii}%
\bibitem{Berezhkovsii&Pollak}%
  \BibitemOpen
  \bibfield{author}{%
  \bibinfo {author} {\bibfnamefont{A.}~\bibnamefont{Berezhkovsii}}, \bibinfo
  {author} {\bibfnamefont{A.}~\bibnamefont{Frishman}},\ and\ \bibinfo {author}
  {\bibfnamefont{E.}~\bibnamefont{Pollak}},\ }%
  \bibfield{journal}{%
  \bibinfo {journal} {J. Chem. Phys.}\ }%
  \textbf{\bibinfo {volume} {101 (6)}},\ \bibinfo {pages} {4778} (\bibinfo
  {year} {1994})%
  \bibAnnoteFile{NoStop}{Berezhkovsii&Pollak}%
\bibitem{Cregg&Wickstead}%
  \BibitemOpen
  \bibfield{author}{%
  \bibinfo {author} {\bibfnamefont{P.~J.}\ \bibnamefont{Cregg}}, \bibinfo
  {author} {\bibfnamefont{D.~S.~F.}\ \bibnamefont{Crothers}},\ and\ \bibinfo
  {author} {\bibfnamefont{A.~W.}\ \bibnamefont{Wickstead}},\ }%
  \bibfield{journal}{%
  \bibinfo {journal} {J.~Appl.~Phys.}\ }%
  \textbf{\bibinfo {volume} {76}},\ \bibinfo {pages} {4900} (\bibinfo {year}
  {1994})%
  \bibAnnoteFile{NoStop}{Cregg&Wickstead}%
\bibitem{Kalmykov&Waldron}%
  \BibitemOpen
  \bibfield{author}{%
  \bibinfo {author} {\bibfnamefont{Y.~P.}\ \bibnamefont{Kalmykov}}, \bibinfo
  {author} {\bibfnamefont{W.~T.}\ \bibnamefont{Coffey}},\ and\ \bibinfo
  {author} {\bibfnamefont{J.~T.}\ \bibnamefont{Waldron}},\ }%
  \bibfield{journal}{%
  \bibinfo {journal} {J.~Chem.~Phys.}\ }%
  \textbf{\bibinfo {volume} {105}},\ \bibinfo {pages} {2112} (\bibinfo {year}
  {1996})%
  \bibAnnoteFile{NoStop}{Kalmykov&Waldron}%
\bibitem{Szabo:JCP}%
  \BibitemOpen
  \bibfield{author}{%
  \bibinfo {author} {\bibfnamefont{A.}~\bibnamefont{Szabo}}, \bibinfo {author}
  {\bibfnamefont{D.}~\bibnamefont{Shoup}}, \bibinfo {author}
  {\bibfnamefont{S.~H.}\ \bibnamefont{Northrup}},\ and\ \bibinfo {author}
  {\bibfnamefont{J.~A.}\ \bibnamefont{McCammon}},\ }%
  \bibfield{journal}{%
  \bibinfo {journal} {J. Chem. Phys.}\ }%
  \textbf{\bibinfo {volume} {77}},\ \bibinfo {pages} {4484} (\bibinfo {year}
  {1982})%
  \bibAnnoteFile{NoStop}{Szabo:JCP}%
\bibitem{Bowman&Noe}%
  \BibitemOpen
  \bibfield{author}{%
  \bibinfo {author} {\bibfnamefont{G.~R.}\ \bibnamefont{Bowman}}, \bibinfo
  {author} {\bibfnamefont{V.~S.}\ \bibnamefont{Pande}},\ and\ \bibinfo {author}
  {\bibfnamefont{F.}~\bibnamefont{Noe}},\ }%
  \bibfield{journal}{%
  \bibinfo {journal} {\textit{An introduction to markov state models and their
  application to long timescale molecular simulation}, Springer: Dortrecht, The
  Netherlands}}%
   (\bibinfo {year} {2014})%
  \bibAnnoteFile{NoStop}{Bowman&Noe}%
\bibitem{Hummer&Szabo}%
  \BibitemOpen
  \bibfield{author}{%
  \bibinfo {author} {\bibfnamefont{G.}~\bibnamefont{Hummer}}\ and\ \bibinfo
  {author} {\bibfnamefont{A.}~\bibnamefont{Szabo}},\ }%
  \bibfield{journal}{%
  \bibinfo {journal} {J. Phys. Chem. B}\ }%
  \textbf{\bibinfo {volume} {119(29)}},\ \bibinfo {pages} {9029} (\bibinfo
  {year} {2015})%
  \bibAnnoteFile{NoStop}{Hummer&Szabo}%
\bibitem{Mondal:JCTC}%
  \BibitemOpen
  \bibfield{author}{%
  \bibinfo {author} {\bibfnamefont{J.}~\bibnamefont{Mondal}}, \bibinfo {author}
  {\bibfnamefont{R.~A.}\ \bibnamefont{Friesner}},\ and\ \bibinfo {author}
  {\bibfnamefont{B.~J.}\ \bibnamefont{Berne}},\ }%
  \bibfield{journal}{%
  \bibinfo {journal} {J.~Chem.~Theory~Comput.}\ }%
  \textbf{\bibinfo {volume} {10}},\ \bibinfo {pages} {5696} (\bibinfo {year}
  {2014})%
  \bibAnnoteFile{NoStop}{Mondal:JCTC}%
\bibitem{Smith:SPCEvisco}%
  \BibitemOpen
  \bibfield{author}{%
  \bibinfo {author} {\bibfnamefont{P.~E.}\ \bibnamefont{Smith}}\ and\ \bibinfo
  {author} {\bibfnamefont{W.~F.}\ \bibnamefont{van Gunsteren}},\ }%
  \bibfield{journal}{%
  \bibinfo {journal} {Chem. Phys. Lett.}\ }%
  \textbf{\bibinfo {volume} {215}},\ \bibinfo {pages} {4} (\bibinfo {year}
  {1993})%
  \bibAnnoteFile{NoStop}{Smith:SPCEvisco}%
\bibitem{Vega:SPCEtens}%
  \BibitemOpen
  \bibfield{author}{%
  \bibinfo {author} {\bibfnamefont{C.}~\bibnamefont{Vega}}\ and\ \bibinfo
  {author} {\bibfnamefont{E.}~\bibnamefont{de~Miquel}},\ }%
  \bibfield{journal}{%
  \bibinfo {journal} {J. Chem. Phys.}\ }%
  \textbf{\bibinfo {volume} {126}},\ \bibinfo {pages} {154707} (\bibinfo {year}
  {2007})%
  \bibAnnoteFile{NoStop}{Vega:SPCEtens}%
\end{thebibliography}%

\end{document}